\documentstyle[eqsecnum,twocolumn,aps,prd,epsfig,times]{revtex}
\tighten
\begin{document}

\def\be{\begin{equation}}
\def\ee{\end{equation}}
\def\bea{\begin{eqnarray}}
\def\eea{\end{eqnarray}}
\wideabs{
\preprint{WISC-MILW-99-TH-07}

\title{Is the squeezing of relic gravitational waves
produced by inflation detectable?}

\author{
Bruce Allen\\
Department of Physics, University of Wisconsin - Milwaukee,
PO Box 413, Milwaukee WI 53201.}

\author{
\'Eanna \'E. Flanagan\\
Newman Laboratory of Nuclear Studies, Cornell University,
Ithaca, NY 14853-5001.}

\author{
Maria Alessandra Papa\\
INFN, Laboratori Nazionali di Frascati, 00040 Frascati (RM), Italy.\\
$\quad$\\
} 

\date{\today}

\maketitle
\begin{abstract}
Grishchuk has shown that the stochastic background of gravitational
waves produced by an inflationary phase in the early Universe has an
unusual property: it is not a stationary Gaussian random process.  Due
to squeezing, the phases of the different waves are correlated in a
deterministic way, arising from the process of parametric amplification
that created them.  The resulting random process is Gaussian but
non-stationary.  This provides a unique signature that could in
principle distinguish a background created by inflation from stationary
stochastic backgrounds created by other types of processes.  We address
the question: could this signature be observed with a gravitational
wave detector?  Sadly, the answer appears to be "no": an experiment
which could distinguish the non-stationary behavior would have to last
approximately the age of the Universe at the time of measurement.  This
rules out direct detection by ground and space based gravitational wave
detectors, but not indirect detections via the electromagnetic Cosmic
Microwave Background Radiation (CMBR).
\end{abstract}
}

\section{Introduction}
\label{s:intro}
The physical processes that took place early in the history of the
Universe are well understood at characteristic times $t\gtrsim 1 \ \rm sec$,
where $t$ denotes proper time after the big bang, in the rest frame of
the cosmological fluid.  However the processes that took place at
times much earlier than this, in what is often called the ``very early
Universe'' are not well understood \cite{KolbTurner}.

A number of theoretical models of the very early Universe have been
constructed which differ significantly from each other.  In many of
these models, phase transitions that take place as the Universe cools
(or supercools) play an important role.  While it can be difficult to
calculate the observational properties of a given model, when this can
be done their parameters can often be chosen so that, at late times,
the resulting cosmology is consistent with the known observational
properties of the present-day Universe.  A significant problem is that
there is often no way to distinguish between these different models
based on strictly objective observational criteria.

One of the most successful models of the very early Universe is the
so-called inflationary model.  Among the many variations of this basic
idea, the most generic and well-studied is the canonical model of
``slow-roll inflation'' \cite{KolbTurner,Linde}.  In this model, the
stress-energy tensor of the very early Universe is dominated by a
vacuum energy term, which results in an extended epoch of exponential
(or near-exponential power-law) expansion.  During this epoch, the
visible part of the Universe becomes extremely homogeneous and smooth,
because the initial fluctuations in the energy density are redshifted
into insignificance.

Inflationary models have been extremely successful for several reasons.
First, they solve two outstanding problems of modern cosmology, the
horizon and flatness problems.  Second, they can be described and
understood with simple analytical models.  Third, they make minimal
assumptions.  And finally, they make very definite observational
predictions about the present-day properties of the Universe.  In
certain cases, for example in their predictions about the temperature
anisotropies in the cosmic background radiation, these inflationary
models are in excellent agreement with the observational data
\cite{RecentReview}.  However other models of the early Universe, such
as the cosmic string model, are also in good agreement with this data
\cite{StringReview}.

One of the ways in which different models of the very early Universe
can (at least in principle) be distinguished is in the predictions
that they make about the stochastic background of gravitational
radiation \cite{LesHouches,allenromano}.  This is a weak background of
gravitational radiation, typically isotropic in its distribution,
which is produced at very early times either by the large-scale motions of
mass and energy in the cosmological fluid, or by the process of parametric
amplification engendered by the expansion of the
Universe.  Because gravitational
forces couple so weakly, this radiation typically evolves free of
disturbing influences, and at the present time, its properties provide
a picture of the state of the Universe at very early times.  To
emphasize this point, consider that the present age of the Universe is
approximately
\begin{equation}
T_0 = {2 \over 3} {H_0}^{-1},
\end{equation}
where the
Hubble expansion rate today is
\begin{eqnarray}
H_0 & = & 100 \> {\rm km \> sec}^{-1} \> {\rm
Mpc}^{-1} \>h_{100} \cr
& = & 3.2 \times 10^{-18} {\rm sec}^{-1} \>h_{100}.
\end{eqnarray}
Here $h_{100} \approx 0.65$ is a dimensionless parameter.  The cosmic
microwave background radiation (CMBR) provides us with a picture of the
structure of 
the Universe at a time about $t \approx 100,000 \>\text{years} \approx
3 \times 10^{12}\> \rm sec$ after the big bang. This should be
compared to the present time $T_0 \approx 3 \times 10^{17} \> \rm
sec$.  On general grounds, the generation of ground-based
gravitational wave detectors currently approaching completion might be
able to detect a relic background of gravitational waves produced
around $t \approx 10^{-22} \> \rm sec$ after the big bang, providing
us with a picture of the very early Universe \cite{LesHouches}.

Of particular interest to us are the two LIGO detectors which will
begin engineering shakedown within the next year \cite{ligo}, and the
European VIRGO \cite{virgo} and GEO-600 projects \cite{geo}.  These
ground-based detectors are sensitive in the frequency range from $\sim 10^1$~Hz
to $\sim 10^3$~Hz; long baseline detectors in space such as the proposed LISA
\cite{lisa} project would extend the frequency range downwards by
about four orders of magnitude to $\sim 10^{-3}$~Hz.  The exciting
prospect is to use these instruments to learn something about the
physical processes that took place in the very early Universe.

Models of the very early Universe typically predict a relic background
of gravitational waves which is stochastic, in the sense that the
gravitational strain at any point in the Universe is a
non-deterministic function of time which can only be characterized in
a probabilistic way.  Often, the gravitational strain arises from the
sum of a large number of independent processes and so the central
limit theorem implies that the resulting random process is
Gaussian.  In addition, in many models the stochastic background is
stationary to a good approximation.

Grishchuk has shown that the stochastic gravitational wave background
produced by inflation is Gaussian, but is {\it not} stationary
\cite{grish,grish89,grishnasa}.  This is because, in inflationary
models, the  
gravitational waves are produced by a process of parametric 
amplification of vacuum fluctuations, resulting in a squeezed quantum
state today.  While Grishchuk's derivation used the language and
formalism of quantum optics, the non-stationarity has also been derived
using the standard methods of curved spacetime quantum field theory
\cite{starobinsky,albrecht0}.

The non-stationarity of relic waves from inflation is significant for
two reasons.  First, one calculation 
indicates that it might make the background easier to detect, because
the integrated signal-to-noise ratio can be made to rise faster than
the square root of the observation time \cite{sqrt}, which is how it
would rise for a stationary and Gaussian background \cite{grish}.
Second, it provides in principle a definite test of inflation: if the
squeezing is observed then it provides 
additional evidence in favor of inflation, and if the background is
observed to be not squeezed, then it falsifies the theory.

In this paper we present a detailed analysis of the detectability of
the non-stationary statistical properties of relic gravitational
waves.  We show that, unfortunately, the effect can not be
observed in an experiment whose duration is short compared to the
age of the Universe at the time of the measurement,
which rules out the possibility of measuring the non-stationarity with 
future ground-based and space-based gravitational wave detectors.
The reason is a combination of two effects.  First, the statistical
properties of the background today vary over a frequency scale $\sim
T_0^{-1} \sim 10^{-17} \, {\rm Hz}$, and any practical experiment
(whose duration is of order a few years or less)
will have a frequency resolution which is coarse compared to
$T_0^{-1}$. 
Second, there is an inherent and unavoidable
limitation in the accuracy of our measurement of statistical
properties of a random process, due to the stochastic nature of the
random process.  This limitation is analogous to what is called
``cosmic variance'' in measurements of CMBR anisotropies.  

We also show that even in a gedanken experiment lasting long enough to
distinguish between the stationary and non-stationary background, there
does not appear to be any way to exploit the latter to make the
integrated signal to noise grow faster than the square root of the
integration time \cite{sqrt}.

We note in passing that it may be possible in the future to probe the
squeezed nature of relic gravitons, albeit indirectly.  Sakharov
oscillations in the angular spectrum of temperature fluctuations in the
CMBR are produced by a combination of tensor perturbations
(gravitational waves) and scalar perturbations.  The existence of these
Sakharov oscillations is directly related to the squeezed nature of the
tensor and scalar perturbations \cite{albrecht1}.  If it is possible in
the future to disentangle the scalar and tensor contributions to the
CMBR anisotropies (perhaps using polarization information \cite{KK}),
it may be possible to confirm that the gravitational waves are indeed
squeezed.  However it appears that will not be possible with the
upcoming MAP and PLANCK experiments \cite{starobinskyaa}.
 Note that the possibility of such a measurement is consistent with our
analysis, since CMBR photons effectively ``measure'' the gravitational
perturbations over a time larger than or of order the age of the
Universe at recombination.

The paper is organized as follows.  
In Appendix \ref{s:appendix_inflation} we review the predictions of
general inflationary models for the statistical properties of relic
gravitational waves.  In Sec.\ \ref{s:simplified2} we extract the
cogent features of those predictions in the context of a simple model
problem, a scalar field in $1+1$ dimensions, and 
describe the differences between {\it stationary} and {\it squeezed}
random processes in this context.
In the remainder of Sec.\ \ref{s:simplified} and in Appendices
\ref{s:appendix_stats} and \ref{s:appendix_stats1}
we analyze in detail measurements that take place at one point in
space.  We show that in order to distinguish 
between the {\it stationary} and {\it squeezed} processes in this context,
a necessary but in general not sufficient condition is that individual
modes need to be resolved with a precision that can only be achieved
with an experiment lasting a substantial fraction of the age of the
Universe.  The arguments do not rely on any specific inflationary
model.   Similar conclusions were reached in work by Polarski and
Starobinsky (page 389 of \cite{starobinsky}).

In Sec.\ \ref{s:inflation} we present a more detailed analysis that
relaxes the assumptions of our simplified analysis of
Sec.~\ref{s:simplified}. 
We consider an explicit model of slow-roll
inflation, and calculate the two-point correlation function
$C(t,t')=\langle h_1(t) h_2(t') \rangle$ of detector strain at two
different sites.  This is the quantity that will be measured in the
upcoming generation of experiments in searches for a stochastic
background \cite{LesHouches,allenromano}.  We are able to obtain and
confirm the main results of Ref.\ \cite{grish}: due to the squeezing,
non-stationary terms (not functions of $t-t'$) arise in $C(t,t')$.
Unfortunately we also show that the non-stationary
terms are averaged away in any observation which is short compared to
$T_0$.  In addition, from the explicit expression of the correlation
function it can be seen that the integrated signal to noise after
optimum filtering should grow with the square root of the integration
time \cite{sqrt} and not faster.

Of course it is a dangerous game to claim that something is not
possible.  The history of physics is full of ``no-go'' theorems that
have been circumvented by clever experiments.  But our work does show
that identifying the squeezed nature of the gravitational waves
produced in an inflationary Universe is probably not possible with
ground and space based gravitational wave detectors.

\section{Simplified Detectability analysis}
\label{s:simplified}

\subsection{Stationary and Squeezed random processes}
\label{s:simplified2}

The statistical properties of relic gravitational waves, as predicted
by inflationary models, are summarized in Appendix \ref{s:appendix_inflation}.
In this section we present an analysis of the detectability of the
non-stationarity in the following simplified context.
Consider a flat $1+1$-dimensional spacetime with topology $R \times
S^1$, with spatial circumference $2 \pi L$, and metric
\be
ds^2 = L^2(-dt^2 + dx^2),
\label{modelmetric}
\ee
where the periodic spatial coordinate $x \in [0,2 \pi)$ and $t$ is now
a dimensionless time coordinate.  
We assume a compact spatial topology for technical convenience only;
the discrete mode normalization is simpler than the corresponding
continuum normalization. 
As a model of gravitational wave perturbations, 
consider solutions to the scalar massless wave equation 
\begin{equation}\label{waveeq}
\left( - {\partial^2 \over \partial t^2 } +  {\partial^2 \over
\partial x^2 } \right) h = 0
\end{equation}
in the
spacetime (\ref{modelmetric}).  The periodic nature of 
the spatial sections means that solutions to the wave equation are
periodic in time\footnote{
Modulo the uninteresting solutions which are linear functions of time
and independent of $x$.  We shall restrict attention to the periodic
solutions, and consequently 
without loss of generality we can also restrict the range of the time
coordinate to $t \in [0,2\pi)$ and imagine that the Universe has
topology $S^1 \times S^1$.}, and have discrete frequencies.  The
frequencies are separated by $\Delta f = 1/(2 \pi L)$.

The analog in this context of a stationary, Gaussian stochastic
background (the naive prediction of inflationary models)
is a scalar field of the form \footnote{The $n=0$ terms have no
gravitational wave analog and should be excluded from all formulae.}:
\begin{equation}
\label{onedstandard}
h_{\rm stationary}(t,x) = \sum_{n=-\infty}^{\infty} r_n \cos( |n| t +
n x + \phi_n ). 
\end{equation}
This solution to the wave equation
(\ref{waveeq}) describes a particular type of stochastic random
process.  It can can be obtained from Eq.\ (\ref{summary1}) below by
dropping the second term, by restricting the unit vector ${\bf n}$
to have two allowed values, and by using a discrete instead of a
continuous mode normalization.
Here the quantities $r_n$ and $\phi_n$ for $n = \pm 1, \pm 2, \ldots$
are random variables whose statistical properties are given by
\be
r_n e^{i \phi_n} = x_n + i y_n,
\label{rphi}
\ee
where $x_n$ and $y_n$ are independent, zero-mean Gaussian random
variables with $\langle x_n^2 \rangle = \langle y_n^2 \rangle =
\sigma_n^2/2$.  The variance $\sigma_n$ can depend on $n$ in an
arbitrary way, but typically in inflationary models $\sigma_n$ has a
power law dependence over a broad range of wavenumbers or
frequencies: $\sigma_n = \alpha |n|^\beta$ for some constants $\alpha$
and $\beta$.  
It follows from Eq.\ (\ref{rphi}) that the $r_n$'s and
$\phi_n$'s are all independent, that the $r_n$'s are Rayleigh
distributed, and that the $\phi_n$'s are uniformly distributed over
the interval $[0,2 \pi)$.

What Grishchuk has shown is that inflation leads to a slightly
different random process, one which to very good approximation can be
written as
\begin{equation}
\label{onedgrishchuk}
h_{\rm squeezed}(t,x) = \sum_{n=-\infty}^{\infty} \sqrt{2} \, r_n \cos
( n t) \cos 
( n x + \phi_n ),
\end{equation}
where the $r_n$'s and $\phi_n$'s are distributed as before.  The factor of
$\sqrt{2}$ ensures that the processes $h_{\rm stationary}$ and $h_{\rm
squeezed}$ have the same time-averaged energy density.
Note that the stationary random process (\ref{onedstandard}) is
invariant under changes $t \to t - \Delta t$ of the origin of the time
coordinate, but that the squeezed random process (\ref{onedgrishchuk})
is not.  In Sec.\ \ref{s:inflation} and Appendix \ref{s:appendix_inflation} 
we show that in inflation models, the simple form (\ref{onedgrishchuk}) is
achieved when one chooses a particular origin for the conformal time
coordinate during the inflationary epoch.  
Again, Eq.\ (\ref{onedgrishchuk}) can be obtained from Eq.\
(\ref{summary1}) below by  restricting the unit vector ${\bf n}$
to have two allowed values, by using a discrete instead of a
continuous mode normalization, by taking $\chi(k)=0$, and by
specializing the the limit of large squeezing $|\beta_k| \gg 1$.

Note that we can rewrite the squeezed random process
(\ref{onedgrishchuk}) as 
\begin{equation}
\label{onedgrishchuk1}
h_{\rm squeezed}(t,x) = \sum_{n=1}^{\infty} \sqrt{2} \, {\bar r}_n
\cos( n t) \cos  
( n x + {\bar \phi}_n ),
\end{equation}
where now the sum is only over positive values of $n$, and 
\be
{\bar r}_n e^{i {\bar \phi}_n} = r_n e^{i \phi_n} + r_{-n} e^{- i
\phi_{-n}}.
\label{barxndef1}
\ee
It follows from Eqs.\ (\ref{rphi}) and (\ref{barxndef1}) that 
\be
{\bar r}_n e^{i {\bar \phi}_n} = {\bar x}_n + i {\bar y}_n
\label{barxndef2}
\ee
where ${\bar x}_n$ and ${\bar y}_n$ are independent, zero-mean
Gaussian random variables with $\langle {\bar x}_n^2 \rangle = \langle
{\bar y}_n^2 \rangle = \sigma_n^2$.  Thus the squeezed random process 
can be described by half as many independent random variables as the
stationary one.  We also note that a solution to the two-dimensional
wave equation is completely 
specified by $h$ and $ \dot h$ at some instant in time, where overdot
denotes $\partial/\partial t$.  Thus, given the two functions
$f(x)=h_{\rm stationary}(0,x)$ and $g(x)=\dot h_{\rm stationary}(0,x)$
one can uniquely determine $r_n$ and $\phi_n$.  The function $h_{\rm
squeezed}$ has only half the number of degrees of freedom, and it is
easy to show that it is a solution to the wave equation determined
entirely
\footnote{It is determined up to an overall multiplicative constant,
except on a set of measure 0 of initial conditions.}  by
$f(x)=h_{\rm squeezed}(0,x)$.

Below we shall be concerned with observations at a fixed point in
space, which without loss of generality we can take to be the point
$x=0$.  The two random processes evaluated at $x=0$ yield
\begin{eqnarray}
\label{distinguish}
h_{\rm stationary}(t,0) & = & \sum_{n=1}^{\infty} \left[ {\hat x}_n
\cos( n t) + {\hat y}_n \sin (n t) \right] \cr 
h_{\rm squeezed}(t,0) & = & \sum_{n=1}^{\infty} \sqrt{2} \, {\bar x}_n
\cos( n t), 
\label{distinguisha}
\end{eqnarray}
where
\begin{eqnarray}
{\hat x} &=& x_n + x_{-n} \nonumber \\
\mbox{} {\hat y}_n &=& - y_n + y_{-n}.
\label{hatxndef}
\end{eqnarray}
From Eqs.\ (\ref{barxndef1}), (\ref{barxndef2}) and (\ref{hatxndef})
the random variables ${\bar x}_n$, ${\bar y}_n$, ${\hat x}_n$ and
${\hat y}_n$ are all independent, zero mean Gaussians with variance
$\sigma_n^2$.

Another convenient way of representing the processes $h_{\rm stationary}(t,0)$
and $h_{\rm squeezed}(t,0)$ is to go from the time domain
representation of these 
fields to their frequency domain representation, through the Fourier
transformation equations. 
The signals of Eqs.~(\ref{distinguish})
can be expressed as a superposition of a discrete infinite set of
cosine functions: 
\begin{equation}
\label{univ_modes}
h(t,0) = \sum_{n=1}^{\infty} R_n \cos( n t + \theta_n ) = {\rm Re}
\sum_{n=1}^{\infty} R_n {\rm e}^{ i(n t + \theta_n)} ,  
\end{equation}
defined, for every frequency $n$, by the Fourier amplitudes $R_n$ and
phases $\theta_n$. These are what we shall refer to as ``the Universe
modes''. The relations between the coefficients appearing in
Eq.~(\ref{univ_modes}) and those in Eq.~(\ref{distinguish}) are, for
the {\it stationary} case:
\begin{eqnarray}
\label{phases}
R_n &  = & \sqrt{ {\hat x}_n^2 + {\hat y}_n^2}
\cr
\theta_n & = & \arg \big[
{\hat x}_n + i {\hat y}_n \big],
\end{eqnarray}
where the range of $\arg$ is $[0,2 \pi)$, and for the {\it squeezed} case:
\begin{eqnarray}
\label{phaseg}
R_n &  = & \sqrt{2} \, |{\bar x}_n| \cr
\theta_n & =  & \cases{
0  ~~~~ {\rm{if}} ~~~~{\bar x}_n > 0\cr
\pi ~~~~{\rm{if}}~~~~ {\bar x}_n < 0}.
\end{eqnarray}

\begin{figure}
\centering
\psfig{file=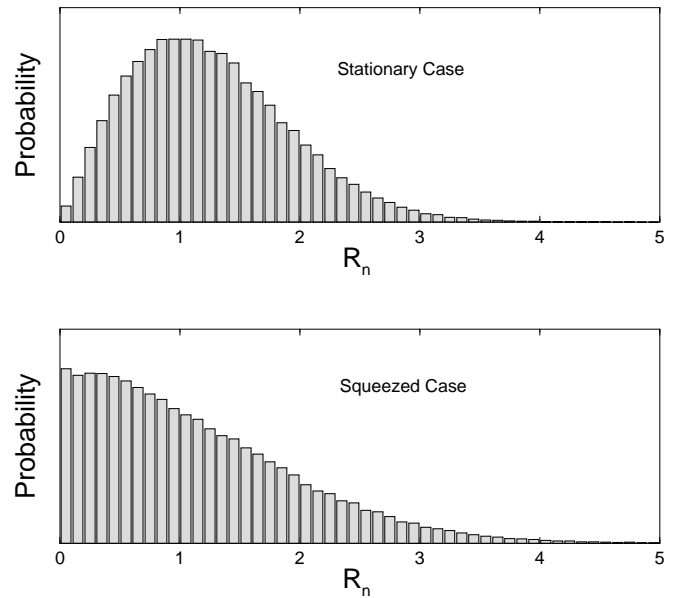,width=8.0cm,angle=-90,clip=}
\caption{Histogram of amplitudes $R_n$ for the {\it stationary} (upper
panel) and 
{\it squeezed} (lower panel) cases for a mode with $\sigma_n^2=1$,
from a total of $10^5$ points.}
\label{hist_Rn}
\end{figure}

\begin{figure}
\centering
\psfig{file=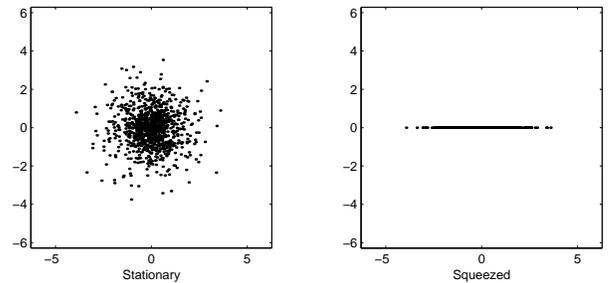,width=8.0cm,clip=}
\caption{The distribution of the complex vectors of a mode  with
$\sigma_n^2=1$ in the {\it stationary} and in the {\it squeezed} case
(from a total of $10^3$ points). }
\label{vect_dist}
\end{figure}

It is convenient to think of each mode as a vector in the complex plane
having length $R_n$ and phase $\theta_n$. This allows one to perform
calculations using complex exponentials - which is easier - and also to
have a useful pictorial representation of what the field is. 
In the {\it stationary} case, it can be seen from Eq.~(\ref{phases})
that the phases $\theta_n$ are uniformly distributed on the interval
$[0,2\pi)$.  In the {\it squeezed} case, instead, the phases
can only have two values, namely $0$ or $\pi$. This means that whereas
the mode vectors for the {\it stationary} case are randomly pointing
in any direction in the complex plane, in the {\it squeezed} case they
all lie on the real axis (see Fig.~\ref{vect_dist}). This is the
origin of the name ``squeezed''. Also the distribution of the
amplitudes $R_n$ is quite different in the two cases, as can be seen
from Fig.~\ref{hist_Rn}.


\subsection{Distinguishing between the two random processes in principle}
\label{s:observational0}

The question that must be addressed is this: what experiment can we
perform that will in practice be able to distinguish between these two
different random processes, (\ref{onedstandard}) and
(\ref{onedgrishchuk})?  In other words, can we construct some
observational statistic that will enable us to determine with high
confidence which of these two random processes is present, given a
single realization?

To appreciate this issue a bit more clearly, consider that both
$h_{\rm stationary}$ and $h_{\rm squeezed}$ are solutions to the
two-dimensional wave equation (\ref{waveeq}).
This fact
can help us to compare the properties of these two random processes.
Roughly speaking, the standard process $h_{\rm stationary}$ is a sum
of left and right moving waves with arbitrary amplitudes and phases:
\begin{eqnarray}
\label{onedstandard1}
h_{\rm stationary}(t,x) & = & \sum_{n=1}^{\infty} \bigg\{ 
r_n \cos[ n (t +  x) + \phi_n ] \nonumber \\
\mbox{} && + r_{-n} \cos[ n (t -  x) + \phi_{-n} ]
\bigg\}. 
\end{eqnarray}
In contrast to this, the Grishchuk process $h_{\rm squeezed}$ is a
sum of standing waves
\begin{eqnarray}
h_{\rm squeezed}(t,x) & = & \sum_{n=1}^{\infty} {{\bar r}_n \over \sqrt{2}} \bigg\{ 
  \cos[ n (t +  x) + {\bar \phi}_n ] 
\nonumber \\
\mbox{} && +  \cos[ n (t -  x) - {\bar \phi}_{n} ] 
\bigg\},
\end{eqnarray}
for which the left and right moving waves have equal amplitudes and
correlated phases.

Now suppose that we ask the following question: could we distinguish
between these two possible forms of $h(t,x)$ based on unrestricted
observations?  The answer is obviously ``yes''.  We would simply
observe the evolution and see if there was a standing wave pattern or not:
a standing-wave pattern would imply $h_{\rm squeezed}$,  
anything else would falsify inflation.

A slightly more difficult question is the following: suppose that
instead we can only observe the function at a single point in space:
$h(t,x=0)$.  Again, could we distinguish between these possibilities?
Once again, the answer is clearly yes.  
By observing the process $h(t,0)$ for $0 \le t < 2 \pi$ and Fourier
transforming, we can recover the complex coefficients $R_n \exp[i
\theta_n]$ in the expansion (\ref{univ_modes}).  If any of these
coefficients have non-zero imaginary parts, then the process must be
stationary rather than squeezed, by Eqs.\ (\ref{phaseg}).

However, our assumption of spatial periodicity (which ensures the
periodicity in time) is just a computational device -- essentially an
infrared cutoff.  We must at the end of our calculation let this
cutoff go to infinity to obtain physical results.  In this limit,
the measurement discussed above corresponds to a measurement of the
random process over an infinite time, since the proper-time duration
of the measurement is $2 \pi L$ which diverges as $L \to \infty$.
So, the conclusion so far is that it would be possible to distinguish
between the {\it stationary} and {\it squeezed} processes, provided
that one had complete knowledge of the function $h(t,x=0)$ over the
entire time history of the Universe.

\subsection{Distinguishing between the two random processes in practice}
\label{s:observational}

Consider now realistic measurements of the stochastic background.
Such measurements will be subject to three key constraints:

\begin{itemize}
\item
The function $h(t,x=0)$ can only be observed over a limited range of
time $T_s \le \tau \le T_s + T_{\rm obs}$, where $\tau = L t$ is
proper time, the starting time $T_s$ is essentially the age of the
Universe $T_0$, and the duration $T_{\rm obs}$ of the
measurement is much smaller than $T_0$. 
In practice the longest possible observation times will be $T_{\rm obs}
\approx 1 \> {\rm year} \> \approx 3 \times 10^7 \> {\rm sec}$.

\item
The detectors available to us can not observe all of the modes, but
just a (fairly narrow) range in frequency $f \in [f_0,f_0+\Delta f]$
where typically  $f_0 \sim \Delta f \sim 100 \> \rm Hz$
for ground based detectors, and $f_0 \sim 10^{-4} \, {\rm Hz}$,
$\Delta f \sim 10^{-2} \, {\rm Hz}$ for space based detectors.

\item
The detectors that measure $h(t,0)$ are intrinsically noisy, limiting
the accuracy of our measurements.
\end{itemize}

\noindent
We now show that the short observation time constraint ($T_{\rm obs}
\ll T_s$ ) alone makes it impossible to distinguish between {\it
stationary} and {\it squeezed}.  
In order to show this, we will derive the relation between the Universe
modes and the observed modes that we actually measure.  We will
show that the latter are a weighted superposition of the former ones,
and that the process of superposition makes it impossible to trace
back the phase coherence of the {\it squeezed} case.

Our analysis in this section will be limited to observations made at a
single point in space.  This assumption is of course not realistic, as 
realistic measurements will involve cross-correlating between
spatially separated detectors.  We present a more complete analysis
which relaxes this restriction in  Sec.\ \ref{s:inflation} below.

We start by making a convenient change in our representation of the
stochastic background.  
So far, we have treated the stochastic background as a continuous
function of $t$ -- a random process $h(t,0)$ -- whose frequency
representation is discrete 
due to the assumed infrared cutoff (spatial periodicity).  We now 
switch to a discrete representation of the stochastic background in
the time domain, by modeling the background as a set of $N$ numbers
\be
h_j = h(j \Delta t),
\label{timedomain}
\ee
for $0 \le j \le N-1$ where $N \Delta t = 2 \pi$.  Thus, we have
effectively imposed an 
ultraviolet cutoff.  At the end of our calculation we can let $\Delta
t \to 0$ and obtain a result that is independent of $\Delta t$ [cf.\
Eq.\ (\ref{ans1}) below].  The discrete
Fourier transform (DFT) of the sequence (\ref{timedomain}) is given by
the relation
\begin{equation}
\tilde{h}_j=\sum_{k=0}^{N-1}e^{2\pi\imath {jk\over N}}h_k,
\label{fourier}
\end{equation}
whose inverse is
\begin{equation}
{h_k}={1\over N}\sum_{j=0}^{N-1}e^{-2\pi\imath {jk\over N}}\tilde{h}_j.
\label{inv_fourier}
\end{equation}
We will also assume that the statistical properties of the quantities
${\tilde h}_j$ are the same as those of the quantities $R_n \exp[i
\theta_n]$ of Eq.\ (\ref{univ_modes}) above, namely
\begin{eqnarray}
\langle ({\rm Re} \, {\tilde h}_j)^2 \rangle &=& (1 + \varepsilon)
\sigma_j^2 \nonumber \\
\mbox{} 
\langle ({\rm Im} \, {\tilde h}_j)^2 \rangle &=& (1 - \varepsilon)
\sigma_j^2,
\label{stats2}
\end{eqnarray}
where $\varepsilon=0$ in the stationary case and $\varepsilon=1$ in
the squeezed case.  Equation (\ref{stats2}) will not be exactly true
but will be true to an adequate approximation.

Lets now take our measurement to consist of the quantities 
\be
H_A = h_{s + A}
\label{measured}
\ee
for $0 \le A \le M-1$, where $s$ and $M$ are fixed integers with $0 <
s < N$ and $0 < M \le N-s$.
Thus, the starting proper-time of the measurement is $T_s = s L \Delta
t$ and 
the duration of the measurement is $T_{\rm obs} = M L \Delta t$.   
Henceforth capital Roman indices ($A$, $B$, $C$, $\ldots$) will be
measurement indices that run over $0,1, \ldots, M-1$, while lower case
Roman indices ($i, j, k, \ldots$) will run over $0, 1, \ldots, N-1$.
The DFT 
\be
{\tilde{H}}_A=\sum_{B=0}^{M-1}e^{2\pi\imath {A B\over M}}H_B
\label{measuredmodes}
\ee
of the quantities (\ref{measured}) define the amplitudes of what we
will call the measured modes.

We can compute the relation between the measured mode amplitudes
(\ref{measuredmodes}) and the Universe mode amplitudes (\ref{fourier})
by 
inserting Eqs.\ (\ref{inv_fourier}) and (\ref{measured}) into Eq.\
(\ref{measuredmodes}), which gives
\begin{equation}
\tilde{H}_A=\sum_{p=0}^{N-1}W_{Ap}\tilde{h}_p.
\label{dft_rel}
\end{equation}
Here the coefficients $W_{Ap}$ are given by
\begin{eqnarray}
\label{wjp0}
W_{Ap} & = &{1 \over N} \, e^{- 2 \pi \imath {p s \over N}} \,
\sum_{B=0}^{M-1} \, e^{2 \pi \imath ( {A \over M} - {p \over N} ) B} \\
\mbox{} &=& 
\displaystyle{1\over N}\ e^{- 2 \pi \imath {p s \over N}} \,
\left[ \displaystyle{ 
1-e^{- 2\pi\imath {pM\over N}}\over 1-e^{2\pi\imath ({A\over M}-{p\over N})}
}\right].
\label{wjp}
\end{eqnarray}

\begin{figure}[t]
\centering
\psfig{file=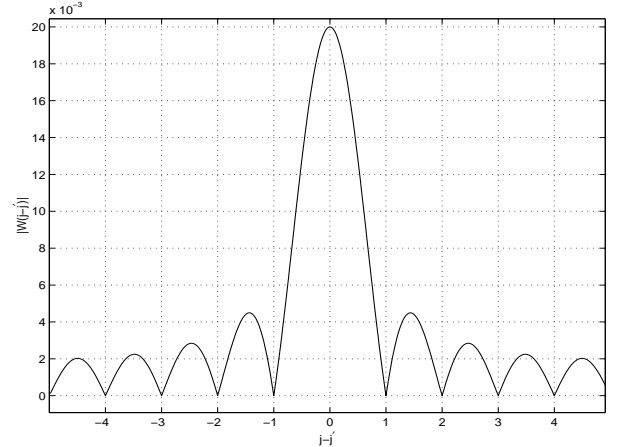,width=8.0cm,height=6.0cm,clip=}
\caption{This plot shows the behavior of $|W_{AA^\prime}|$ as 
a function of $A-A^\prime$ for the case $M=10$, $N = 500$. The maximum
value is ${M\over N}={10\over 500}$.} 
\label{wjjp_j5}
\end{figure}

Equations (\ref{dft_rel}) and (\ref{wjp}) say that the complex vectors
$\tilde{H}_A$ (amplitudes of the observed modes) that define the DFT
from the shorter time baseline (\ref{measured}) are a weighted sum of
the vectors $\tilde{h}_p$ (amplitudes of the Universe modes) that
define the DFT from the longer time baseline (\ref{timedomain}).  The
weights are complex numbers whose magnitudes depend only on the
distance in frequency between $A$ and $p$.  In order to make this
statement clearer, let us take a closer look at $W_{Ap}$ and introduce
a new index variable
\be
A^\prime=A{M\over N}.
\ee
This variable can take values from $0$ to $M-1$, as does the index
$A$, but is not an integer.  It measures the position of the points
$p$ on the $A$ axis.  In terms of this new variable the weights take
the form 
\begin{equation}
W_{A A^\prime}=\displaystyle{1\over N} \ 
e^{- 2 \imath\pi A^\prime s / M} \ 
e^{\imath\pi (A-A^\prime)(1 - 1/M)}\  
{ \sin \left[ \pi (A - A^\prime) \right] 
\over \sin \left[ {\pi \over M} (A - A^\prime) \right] } 
\label{wjjprime}
\end{equation}
Note that when $A^\prime$ is an integer, we have $W_{AA^\prime} =
W_{A^\prime A}^*$, which is necessary since the random sequences $h_j$
and $H_A$ are real.  Also we can write
\be
W_{AA^\prime} = {M \over N} f_M(|A - A^\prime|) e^{i
\varphi_{AA^\prime}}
\ee
with
\be
f_M(x)=\displaystyle{\sin{\pi x }\over M 
\sin{{\pi\over M} x }},
\ee
which shows that the absolute values of the weights depend only on the
distance $(A-A^\prime)$ between $A$ and $p$.  
Figure \ref{wjjp_j5} shows the magnitudes $|W_{AA^\prime}|$ of the
weights for the case $M=10$, $N=100$.  This figure 
makes it clear that the Universe modes that contribute the most to a
given observed mode, in the sum (\ref{dft_rel}), are those which are
nearest to it.  It also shows that for any measured mode $A$,
the amplitudes of the Universe modes to the right of $A$
(i.e. modes with $A^\prime > A$) are weighted in the same way as the
modes to its left ($A^\prime < A$), which follows from
$|W_{A\,A+\Delta A}| = |W_{A\,A-\Delta A}|$.

The crucial property of the weighting factors $W_{AA^\prime}$ is that
they contain oscillations on two different frequency scales.  First,
there is above-discussed oscillation in the magnitude $|W_{AA^\prime}|$
with characteristic frequency scale $\sim T_{\rm obs}^{-1}$.  Second,
there is in addition a variation in the phase of $W_{AA^\prime}$,
encoded in the factor $\exp[ -2 \pi \imath A^\prime s / M]$ in Eq.\
(\ref{wjjprime}), with characteristic frequency scale $ \sim T_s^{-1}$.

We can now give the key argument.  Consider a particular measured mode
amplitude ${\tilde H}_A$, which corresponds to a physical frequency
$f_A = A / (M L \Delta t)$.  This mode amplitude results from a
vectorial sum (\ref{dft_rel}) of Universe mode amplitudes ${\tilde h}_j$.
The Universe modes ${\tilde h}_{j = N A^\prime / M}$ which contribute
significantly to ${\tilde H}_A$ all 
lie with a frequency interval around $f_A$ of width $\sim T_{\rm
obs}^{-1}$, because of the oscillations in $|W_{AA^\prime}|$.  Now,
{\it if} the weighting factors $W_{AA^\prime}$ were all real in this
frequency interval, then since in the squeezed case the Universe mode
amplitudes ${\tilde h}_j$ are all real, then the resulting measured
mode amplitude ${\tilde H}_A$ obtained from the sum (\ref{dft_rel})
would also be real.  Hence, the observed mode amplitude would share
with the original Universe mode amplitudes that property that its
phase is constrained to be $0$ or $\pi$, and so the
squeezing would be easily observable.

Thus, we see that the crucial point is the extent to which the weighting
factors can be treated as real in the frequency interval of width
$\sim T_{\rm obs}^{-1}$ around the central frequency $f_A$.  Consider first
the short-measurement regime $T_{\rm obs} \ll T_s$.  In this regime,
the phase of $W_{AA^\prime}$ winds around between $0$ and $2 \pi$
roughly $T_s / T_{\rm obs} \gg 1$ times in the frequency interval, due to
the factor $\exp[ -2 \pi \imath A^\prime s / M]$ in Eq.\
(\ref{wjjprime}), and so cannot be treated as real.  
In this case the observed mode ${\tilde H}_A$ has a phase that is very
nearly uniformly distributed over $[0,2 \pi)$; the vectorial sum
(\ref{dft_rel}) almost completely erases the peculiar phase behavior
of the non-stationary process.  This is in part due to the fact that the 
vectorial sum entangles phase and amplitude information and that the
amplitudes of the Universe modes (although not the phases) are
stochastic.  In Appendix \ref{s:appendix_stats} below we demonstrate this
by deriving the distributions of the phase and amplitude of the
observed amplitude ${\tilde H}_A$.

So far, we have argued that the effect of the squeezing on the
statistical properties of each observed mode ${\tilde H}_A$ is very
small in the regime $T_{\rm obs} \ll T_s$.  This gives the essential
reason why the squeezing is not observable.  However, we still need to
close two loopholes in the argument.  First, 
we have not yet shown that the effect of squeezing is unobservable,
as we have not 
addressed the question of how accurately the statistical properties of
each mode can be measured.  In the case of the CMBR, for example, the
anisotropies are smaller than the homogeneous background CMBR by a
factor of $10^5$, yet those anisotropies are still observable and
carry a great deal of information.  We need to show that the 
intrinsic limitations in the possible accuracy of measurement of the
statistical properties of each mode (``cosmic variance'') are larger
than the effect of squeezing on each mode.  Second, it is
insufficient to perform an analysis that focuses on each individual
measured mode ${\tilde H}_A$ one at a time, as the individual measured
modes are not statistically independent, and we need to show that
squeezing cannot be observed in any measurement that combines
information from all the measured modes \cite{LIGOeg}.
These two concerns are
addressed and resolved in Appendix \ref{s:appendix_stats}, where we
demonstrate that the 
effect of squeezing is indeed not observable.

For completeness, we now consider the 
long measurement regime $T_{\rm obs} \gg T_s$, although as explained in
the Introduction this regime cannot be realized in practice.
In this regime, the oscillation in the phase of $W_{AA^\prime}$ due to
the factor $\exp[ -2 \pi \imath A^\prime s / M]$ in Eq.\ 
(\ref{wjjprime}) is negligible.  There is a phase oscillation due to
the factor $\exp[ \pi \imath (A - A^\prime)(1 - 1/M)]$ in Eq.\
(\ref{wjjprime}), which gives a total change of phase of order
unity over the frequency interval.  The resulting observed mode amplitude
${\tilde H}_A$ is thus not constrained to be real.  It turns out that
the probability distribution for its phase ${\rm Arg} {\tilde H}_A$ is 
exactly uniform in the case when the spectrum is white, i.e.,
when $\sigma_j = $ const.  When the spectrum is colored, the
detectability of the squeezing depends on how large $T_s$ is compared
to a correlation time determined by the spectrum $\sigma_j$ (see
Appendix \ref{s:appendix_stats}).

\section{Two-detector correlation experiments}
\label{s:inflation}
\subsection{Introduction}

In this section we extend the analysis of Sec.\ \ref{s:simplified} of
the detectability of the non-stationarity to incorporate a number of
complicating effects, including the effect of having measurements at
more than one point in space.  We restrict attention in this section
to the realistic case of short observations, $T_{\rm obs} \ll T_s$.

The standard method which will be used to detect a stochastic
gravitational-wave background is based on correlating two
widely-separated detectors.  In the standard treatments of this
technique, one considers a stochastic background which is
stationary and Gaussian.  In this case it is easy to show that the
integrated two-detector correlation arising from the stochastic
background is proportional to the integration time, whereas the terms
arising from the (assumed uncorrelated) detector noise are
proportional to the square root \cite{sqrt} of the integration time.

As stressed by Grishchuk, the stochastic background produced by an
inflationary epoch in the early Universe is in a squeezed quantum state
produced by the period of rapid expansion.  The statistical properties
of this state are Gaussian, but {\it not} stationary.  
Unfortunately the non-stationary behavior does not make the stochastic
background produced by inflation easier to observe than a similar
background which is stationary and Gaussian\footnote{Recent work by Turner
\cite{turnergw} has shown that within the context of slow-roll
inflationary models, current observational limits on the CMBR
imply that the gravitational wave stochastic
background of inflationary models would be too weak to detect with
either the current LIGO-I or planned upgrade LIGO-II detector, and is
just below the limits of sensitivity of the proposed LISA \cite{lisa}
experiment.}.  We show this in Sec.\ \ref{twopoint} below by deriving an
expression for the correlation function for the detector strains at two
separated sites.  We show that, in contrast with the correlation
function for a stationary Gaussian background (which depends only upon
$t-t'$), the squeezed nature of the inflationary background gives rise
to an extra term in the correlation function.  This extra term
explicitly manifests the non-stationary nature of the squeezed state:
it is {\it not} a function of $t-t'$ alone.  Thus Grishchuk is entirely
correct that the squeezed quantum state has different properties than a
stationary and Gaussian background.  However we show that observing
this extra non-stationary term requires a frequency
resolution of order $10^{-17}$~Hz.  Such incredible frequency
resolution is only obtainable in a gedanken experiment lasting as long
as the present-day age of the Universe.

The calculation in this Section makes frequent use of results from two
references.  The first reference examines the gravitational waves
produced by an inflationary epoch in the early Universe \cite{AK94},
and calculates the correlation function of the temperature
anisotropies that they induce in the 
CBR. The frequency range of the waves influencing the observable
part of the CBR anisotropy is between $10^{-18}\> \text{Hz}$ and
$10^{-16}\> \text{Hz}$ today.  But the methods of Ref.~\cite{AK94} can
also be applied (as we do here) to calculate the properties of the
waves at much higher frequencies: the $10^{-3} \> \text{Hz}$ relevant
to space-based or the $10^{2} \> \text{Hz}$ relevant to earth-based
gravitational-wave detectors.  The second reference that we make use
of is a review article on how correlation techniques may be used to
detect a stochastic background \cite{allenromano}.  A reader wishing
to follow our calculations in detail is advised to have copies of
these two papers in hand.

The calculation done in this Section is for an explicit model of
slow-roll inflation, in the limit that the resulting spectrum is
``flat'' and not tilted.  After completing the explicit calculation of
the correlation function, it is straightforward to show that in more
general inflationary models, which produce a ``tilted'' spectrum, the
same general conclusion holds: the non-stationary behavior can not be
observed in any practical experiment.

\subsection{Calculation of the correlation function $C(t,t')$}
\label{twopoint}

The most direct way to understand how the squeezing produced in an
inflationary model affects the correlation between separated detectors
is to examine the two-point correlation function for the gravitational
wave strain $h$ at two detector sites.  This is defined by:
\be
C(t,t') = \langle h_1(t) h_2(t') \rangle
\label{correlfndef}
\ee
where the subscripts $1$ and $2$ refer to the detector sites, and $t$
and $t'$ denote the time at these sites.  The angle brackets can be
given two possible (equivalent) meanings.  They can denote the average
over a statistical ensemble of many different inflationary Universes,
each starting from somewhat different initial conditions, but made
statistically similar by a long period of exponential inflation.
Equivalently, they can denote the expectation value in a quantum state
that is a good approximation to the present-day state of the
Universe.

The detector strains are given in terms of the metric perturbations of
a Friedman-Robertson-Walker (FRW) cosmological model.  Denoting the
space-time metric by
\begin{equation}
\label{e:metric}
ds^2 = a^2(\eta) \left( -d\eta^2 + (\delta_{ab} + h_{ab}) dx^a dx^b \right)
\end{equation}
the strain at site $i$ is given by:
\be
h_i(t) = {1 \over 2} h_{ab}(t,{\bf x}_i) (\hat X_i^a \hat X_i^b - \hat
Y_i^a \hat Y_i^b)
\label{straindef}
\ee
where ${\bf x}_i$ is the spatial location of the $i$-th site and $\hat
X_i^a$ and $\hat Y_i^a$ are unit-length spatial vectors (with respect
to the metric $\delta_{ab}$) along the directions of the orthogonal
detector arms at site $i$.  Our model of the Universe is specified via
the cosmological scale factor $a(\eta)$ given as a function of
conformal time $\eta$ by Eq.~(\ref{e:scalef}).  Note that we frequently
use cosmological time $t$ rather than conformal time $\eta$ as a
coordinate; they are related by $dt = a(\eta) d\eta$.

To compute the correlation function, we need an expression for the
metric perturbation $h_{ab}$. This quantity is not deterministic.  In
classical calculations, it may be treated as a real stochastic random
variable with certain statistical properties.  For inflationary
cosmological models, a slightly different approach is useful, in which
$h_{ab}$ is replaced by the field operator $\bar h_{ab}$ of the
linearized gravitational field.  To see why this is both simple and
useful, it is helpful to briefly review the fundamental mechanism through
which inflationary models erase the effects of the conditions present
before the inflation begins and give rise to a spatially flat and
homogeneous Universe today.

In inflationary models, the Universe undergoes a period of (nearly)
exponential expansion of the scale factor $a$.  During this period,
the energy-density of any matter or radiation is red-shifted away
exponentially quickly.  For massless particles, the energy of a
particle is proportional to $a^{-1}$ and hence the energy-density in
massless particles redshifts as $a^{-4}$.  For massive particles, any
kinetic energy is quickly red-shifted away, leaving only the rest-mass
behind.  This energy is diluted by the expansion of the spatial
three-volume, so that the energy-density of massive particles
redshifts away as $a^{-3}$.  What is left behind after the period of
exponential inflation is just the vacuum energy associated with the
cosmological constant; it is the dominant form of energy at the end of
an inflationary epoch.

The perturbations away from absolute
uniformity in inflationary models are dominated by the effects of the
zero-point vacuum fluctuations of the fields at the start of inflation
\cite{explain}. 
Twenty five years ago Grishchuk showed that
during the subsequent expansion of the Universe, the quantum
zero-point fluctuations of the gravitational field modes are
parametrically amplified \cite{jurassic_era}.  
This prediction has been subsequently confirmed in many
different ways, and the process of parametric amplification, which
applies to both scalar and tensor perturbations, is one explanation of
how the early Universe can generate a spectrum of perturbations.

The above discussion, combined with Eqs.\ (\ref{correlfndef}) and
(\ref{straindef}), 
implies that the correlation function today is well
approximated by the vacuum expectation value
\begin{eqnarray}
\label{e:corrfun}
C(t,t') & = &{1 \over 8} \langle 0 |
 {\bar h}_{ab}(t,{\bf x}_1) {\bar h}_{cd}(t',{\bf x}_2)
+ 
{\bar h}_{cd}(t',{\bf x}_2)  {\bar h}_{ab}(t,{\bf x}_1) 
 | 0 \rangle \cr
& & (\hat X_1^a \hat X_1^b - \hat Y_1^a \hat Y_1^b)
(\hat X_2^c \hat X_2^d - \hat Y_2^c \hat Y_2^d),
\end{eqnarray}
where $|0 \rangle$ is the (unique Hadamard de Sitter-invariant)
quantum vacuum state of the inflationary Universe \cite{allen89}, and
$\bar h_{ab}$ is the Hermitian field operator of the linearized
gravitational field.

The correlation function (\ref{e:corrfun}) can be calculated exactly,
but since space and ground based detectors will only be sensitive to
gravitational waves in the frequency range $f \in [10^{-3}, 10^{3}]$
Hz, a number of simplifying approximations can be made.
We do this by approximating the field operator within this frequency range.
In general, the field operator is given by Ref.~\cite{AK94} as
\begin{eqnarray}
\bar h_{ab}(\eta,
{\bf x}) &=&
\int d^3k\>  \bigg(e^{i {\bf k} \cdot {\bf x}} \big[e_{ab}( {\bf k} )
\phi_R(\eta,k) \bar a_R( {\bf k}) \nonumber \\
\mbox{} && 
+e^*_{ab}({\bf k})\phi_L(\eta,k)\bar a_L( {\bf k}) \big]
+ {\rm h.c.} \bigg),
\label{e:barhab}
\end{eqnarray}
where ${\rm h.c.}$ means Hermitian conjugate.
This is a sum over wavevectors (labeled by $\bf k$) of left- and
right-circular polarization tensors $e_{ab}$ and $e^*_{ab}$ and
wavefunctions $\phi_{L,R} (\eta,k) \exp( \imath \bf k \cdot \bf x) $.
To approximate the field operator we need to relate present-day
frequency $f$ to $k=|{\bf k}|=\sqrt{k_a k^a}$.  Since $\bf k$ always
appears in the form $\exp( \imath \bf k \cdot \bf x)$, from the form
of the metric (\ref{e:metric}) one can see that the frequency $f$
today is related to $k$ by
\begin{equation}
\label{e:ftok}
f = {k \over 2 \pi a(\eta_0)}.
\end{equation}
We will now express this relationship between $f$ and $k$ in terms of
the present-day Hubble expansion rate $H_0$, which will yield Eq.\
(\ref{k_of_f}).  This leads 
to relations (\ref{ketas}) needed to approximate the field operator
(\ref{e:barhab}) and to evaluate the two-point correlation function
(\ref{e:corrfun}) in the frequency range of interest.

The inflationary cosmological model is defined by the scale factor
$a(\eta)$ in Eq.~(4.1) of Ref.~\cite{AK94}:
\begin{equation}
\label{e:scalef}
a(\eta)=\left\{\begin{array}{lll} \big(2-{\eta\over{\eta_1}}\big)^{-1}
a({\eta_1}) & -\infty<\eta<{\eta_1} &{\rm de\ Sitter,}\\ &&\\
{\eta\over{\eta_1}} a({\eta_1}) & {\eta_1}<\eta<{\eta_2}  & {\rm
radiation,}\\ &&\\ {1\over 4}\big(1+{\eta\over{\eta_2} }\big)^2
{{\eta_2} \over{\eta_1}} a({\eta_1}) & {\eta_2} <\eta &{\rm matter}.
\end{array} \right.
\label{scalefactor}
\end{equation}
This scale factor describes three epochs: a de Sitter inflationary
phase, followed by a radiation-dominated and then a matter-dominated
phase.  Letting $t$ be the cosmological time defined by $dt = a(\eta)
d\eta$, the Hubble constant today is given by:
\begin{eqnarray}
H_0 & = & \left( {1 \over a}  {da \over dt}\right)_{\rm today}
= \left( {1 \over a^2}  {da \over d\eta} \right)_{\rm today} \cr
\label{e:hub1}
& = & {2 \over \eta_0 + \eta_2 } a^{-1}(\eta_0).
\end{eqnarray}
Here $\eta_0$ is the present-day value of the conformal time
and $\eta_2$ is the value of the conformal time at the beginning of
the matter-dominated phase of expansion.  The redshift $Z_{\rm eq}$ at which the
matter and radiation energy densities are equal is
\begin{equation}
\label{e:zeq}
1+Z_{\rm eq} =
{a(\eta_0) \over a(\eta_2)}=
{1 \over 4} (\eta_0/\eta_2 + 1)^2 \approx 10^4.
\end{equation}
Hence $\eta_0$ is about two orders of magnitude greater than $\eta_2$
and the Hubble constant today (\ref{e:hub1}) is well approximated by
\begin{equation}
\label{e:approxhub}
H_0 \approx {2 \over \eta_0 } a^{-1}(\eta_0) \approx 10^{-18} \> \text{Hz}.
\end{equation}
Having specified the cosmological model, one may now approximate the
field operator $\bar h_{ab}$.

We will be making approximations valid for the range of frequencies
that might be observed by earth- or space-based detectors.  It follows
from Eqs.~(\ref{e:ftok}) and (\ref{e:approxhub}) that $k$ is related
to the present-day frequency $f$ by
\begin{equation}
\label{k_of_f}
k \approx {4 \pi f \over \eta_0 H_0 }.
\end{equation}
In the frequency range of interest, $k \eta_0 \approx 4 \pi f/H_0
\gtrsim 10^{16}$ is much larger than unity: $k \eta_0 >> 1$.  From
examination of Eq.\ (\ref{e:zeq}) this implies that one has $k \eta_2
\gtrsim 10^{14}$ so $k \eta_2 >> 1$. Finally, from
Eqs.~(4.2) and (4.3) of Ref.\ \cite{AK94}, if the period of inflation 
creates sufficient 
cosmological expansion to solve the horizon and flatness problems, then
\begin{equation}
{\eta_1 \over \eta_2} = {1 + Z_{\rm eq} \over 1 + Z_{\rm end}} \lesssim
10^{-23}.
\end{equation}
Here $Z_{\rm end} \gtrsim 10^{27}$ is the redshift at the end of the
inflationary epoch.  Hence $k \eta_1 \lesssim 10^{-9}$ is much smaller
than unity: $k \eta_1 << 1$.  
To summarize, we will approximate the field operator in the frequency
regime where
\begin{eqnarray}
\label{ketas}
k \eta_0 & >> & 1, \cr
k \eta_1 & << & 1, \text{ and} \cr
k \eta_2 & >> & 1.
\end{eqnarray}
A physical interpretation of these constraints will be given shortly.

Any present-day detector correlation experiment will observe the
correlation function $C(t,t')$ at the present time: $t$ and $t'$ are
in the matter-dominated (present-day) epoch.  From the definition of
the correlation function (\ref{e:corrfun}) and of the field operator
(\ref{e:barhab}) this means that we need an expression for the mode
function $\phi(\eta,k)$ in the present-day (matter-dominated) epoch,
which is given by the final line of Eq.~(4.17) of \cite{AK94}:
\[
\phi(\eta,k) = \phi_{L,R}(\eta,k) 
= \alpha\>\phi^{(+)}_{\rm mat}(\eta,k) +\beta\>\phi^{(-)}_{\rm
mat}(\eta,k).
\]
The mode functions in the present-day matter-dominated phase are
given by Eq.~(4.7) of \cite{AK94} 
\begin{eqnarray}
\label{e:exactmode}
\phi^{(+)}_{\rm mat}(\eta,k) &=& \phi^{(-)^*}_{\rm mat}(\eta,k)
\nonumber \\
\mbox{} &=& -4 i
 \sqrt{ {8 \over 3 \pi} {\rho_{ds} \over \rho_P} } k^{3/2} \eta_1^2
 \eta_2 { h_1^{(2)}(k(\eta+\eta_2)) \over k (\eta+ \eta_2)},
\end{eqnarray}
in terms of a spherical Hankel function of the second kind
$h_1^{(2)}$.  The Planck density is denoted by $\rho_P = {c^7 \over
\hbar G^2}$.

The Bogoliubov coefficients $\alpha$ and $\beta$
are given in terms of the corresponding coefficients for the transition
between the de Sitter- and radiation-dominated phases and the
transition
between the radiation- and matter-dominated phases as
\begin{equation}
\pmatrix{\alpha &\beta\cr \beta^* &\alpha^*}=
\pmatrix{\alpha &\beta\cr \beta^* &\alpha^*}_{\rm rad} \pmatrix{\alpha
&\beta\cr \beta^* &\alpha^*}_{\rm mat}.  
\label{alphabetadefinition}
\end{equation}
Since $k \eta_1 << 1$ and $k \eta_2 >> 1$, one may make use of
Eqs.~(4.12) and (4.14) of Ref.~\cite{AK94} to obtain:
\be
\alpha \approx - \beta^* \approx {{\rm e}^{\imath k (\eta_1 + \eta_2)}
\over 2 k^2 \eta_1^2}.
\label{e:bogc}
\ee
Finally, making use of the definition of $\phi^{(+)}_{\rm mat}$ given
in Eq.~(\ref{e:exactmode}), one obtains:
\begin{eqnarray}
\phi(\eta,k) & \approx  &  4 \imath k^{-5/2} \sqrt{
{8 \over 3 \pi} {\rho_{ds} \over
\rho_P} } \eta_2 (\eta_0 + \eta_2)^{-2} \cos k(\eta - \eta_1) \cr
& \approx & 
4 \imath k^{-5/2} \sqrt{
{8 \over 3 \pi} {\rho_{ds} \over
\rho_P} } \eta_2 \eta_0^{-2} \cos k\eta
\label{e:finalmode}
\end{eqnarray}
where $\rho_{ds}$ is the constant energy density during the inflationary
de Sitter phase of expansion and $\rho_P$ is the
Planck energy density.  In deriving this expression we have made two
approximations.  
During any period of observation lasting
only a few years
$(\eta+ \eta_2) \approx (\eta_0+ \eta_2)$.
Also 
\[ h_1^{(2)}(x) = -{{\rm e}^{-ix} \over x} (1 - {i \over x}) \approx
-{{\rm e}^{-ix} \over x}
\]
since $x=k(\eta +\eta_2) \approx k(\eta_0 +\eta_2) >> 1$.

Note that the field operator (\ref{e:barhab}) is invariant under the
combined transformations
\begin{eqnarray}
{\bar a}_R({\bf k}) &\to& e^{- i \psi(k)} {\bar a}_R({\bf k}), 
\nonumber \\
\mbox{} {\bar a}_L({\bf k}) &\to& e^{- i \psi(k)} {\bar a}_L({\bf k}), 
\nonumber \\
\mbox{} \phi(k,\eta) & \to & e^{i \psi(k)} \phi(k,\eta),
\end{eqnarray}
where $\psi(k)$ is an arbitrary function of $k$.
The de Sitter vacuum state is invariant under this transformation,
since it is defined by ${\bar a}_R({\bf k}) | 0 \rangle = 
{\bar a}_L({\bf k}) | 0 \rangle = 0$.  Hence without loss of
generality we may multiply the mode function (\ref{e:finalmode}) by a
$k$-dependent phase; the final physical results will not be affected by
such a transformation.  Changing this phase is analogous to multiplying
the wavefunction of a quantum mechanical harmonic oscillator by a pure
phase: it has no observable effects.  In particular, changing the phase
$\psi$ is {\it not} equivalent to changing in the argument of the
$\cos$ appearing in Eq.~(\ref{e:finalmode}).

The expressions for the Bogoliubov coefficients (\ref{e:bogc}) and
the mode functions (\ref{e:finalmode}) have a number of interesting
properties:
\begin{itemize}
\item
The quantity $|\beta |^2$ is the (very large) number of quanta created
by the ``external'' large-scale expansion of the Universe (or
equivalently, by the parametric amplification of zero-point
fluctuations) \cite{birreldavies}.  This is how inflation gives rise
to a potentially-observable stochastic background of gravitational
waves today \cite{allen89}.
\item
The gravitons are created in particle-antiparticle pairs.  Since the
antiparticle of a graviton is just a graviton of opposite momentum and
helicity, gravitons are always created in oppositely-moving pairs
\cite{parker,parker2}.
\item
Since the amplitudes of these oppositely moving pairs of gravitons are
exactly equal and their momenta are opposite \cite{parker,parker2}, they give
rise to a pattern of standing waves.
\item
This pattern of standing waves is apparent in the form of the mode
function $\phi$, whose complex phase is {\it not} a function of time
(since $\phi$ is pure imaginary).
\end{itemize}
These are precisely the conclusions reached
by Grishchuk following Eq.~(7) of reference
\cite{grish}.

The argument $k(\eta - \eta_1)$ of the cosine in Eq.\
(\ref{e:finalmode}) has a simple physical
interpretation.  The number of cycles $dN$ of a wave in the time interval
$dt$ at time $t$ is $dN = f(t)dt = f(\eta) a(\eta) d\eta = k d\eta / 2
\pi$.  This means that $k(\eta - \eta_1)/2\pi$ is the number of cycles
of oscillation (in time) that the wave with wavenumber $\bf k$ has
undergone since the end of the de Sitter phase at time $\eta_1$.  For
frequencies observable by earth- and space-based detectors, the term
$k \eta_1 << 1$ and can be neglected.  See also Appendix
\ref{s:appendix_inflation} below.  Later in this Section, we will
encounter terms of the form $k(\eta \pm \eta')$.  

Before completing the calculation of the two-point correlation
function, it is useful to make a short digression.  We will calculate
the energy-density in gravitational waves using this formalism.  The
result illustrates precisely the effect predicted in
Section~\ref{s:observational}: it is not possible to distinguish the
the {\it stationary} and {\it squeezed} states in local short-time
observations.  The energy-density in gravitational waves is given by
\bea \nonumber \rho_{gw} & = & \int df {d\rho_{gw} \over df} \cr & = &
\lim_{t \to t'} {1 \over 32 \pi G} \langle 0 | {d \over dt} {\bar
h}_{ab}(\eta,{\bf x}) {d \over dt'} {\bar h}^{ab }(\eta ',{\bf x}) | 0
\rangle \cr & = & \lim_{\eta \to \eta'} {1 \over 32 \pi G}{1 \over
a(\eta)a(\eta')}{d \over d\eta}{d \over d\eta '} \langle 0 | {\bar
h}_{ab}(\eta,{\bf x}) {\bar h}^{ab }(\eta',{\bf x}) | 0 \rangle.  
\eea
The last term of the expression above, which is the two-point function
of the field operator, may be derived from the plane-wave
expansion of the field operator (\ref{e:barhab}) and the canonical
commutation relations given by Eq.~(2.17) of \cite{AK94}. Doing so
yields 
\begin{eqnarray}
\label{e:twopointfn}
\langle 0 | {\bar h}_{ab}(\eta,{\bf x}) &&
    {\bar h}_{cd }(\eta',{\bf x'}) 
   | 0 \rangle =  
   \int d^3 k ~ {\rm e}^{\imath {\bf k} \cdot ({\bf x} - {\bf x}') }
   \phi(\eta,k) \phi^*(\eta',k) \nonumber \\
&& \times \left[ e_{ab}({\bf k}) e^*_{cd }({\bf k})  +   e^*_{ab}({\bf k}) e_{cd
   }({\bf k}) \right], 
\end{eqnarray}
since the vacuum state $|0  \rangle$ is annihilated by the operators $
{\bar a}_R$ and ${\bar a}_L$. Note that this quantity also appears in the
definition of 
$C(t,t')$ and will be referred to later in this section.
Making use of the relationship between cosmological and conformal time
$dt = a(\eta) d\eta$ and of the definitions (2.22-26) of \cite{AK94},
one obtains an expression for the energy-density in the stochastic
gravitational wave background valid for the range of frequencies
observable by earth- and space-based detectors:
\[
\rho_{gw} = {64 \over 3 \pi} {\rho_{ds} \over \rho_P} 
{\eta_2^2 \over \eta_0^4 a(\eta_0)^2 } 
\int k^{-1} \cos^2 k \eta ~ dk.
\]
It is conventional to express this energy-density as a dimensionless
spectral function $\Omega_{\rm gw}(f)$ which is the ratio of the
energy-density in gravitational waves in a logarithmic frequency
interval divided by the critical energy-density $\rho_c = {3 \over 8
\pi} H_0^2$ required to close the Universe\footnote{
Many of the ``standard" treatments of gravitational wave production by
slow-roll inflation would imply that the energy-density in
gravitational waves is zero.  This is a misapplication of the standard
consistency relation between the scalar and tensor amplitudes, because
in this non-tilted model the standard treatments imply that the
energy-density in scalar perturbations is
infinite.
}:
\bea
\Omega_{\rm gw}(f) & = & {f \over \rho_c} {d \rho_{gw} \over df} = {32 \over 9}
{\rho_{ds} \over \rho_P} (1 + Z_{eq})^{-1} \cos^2 k\eta \cr
& \approx & {32 \over 9}
{\rho_{ds} \over \rho_P} (1 + Z_{eq})^{-1} \cos^2 [2 \pi f(t+2 T_0)].
\label{e:infomega}
\eea
Note that $\Omega_{\rm gw}(f)$ depends upon the time: if the
statistical properties of the state were stationary, it would
be time-independent.  For a single mode, the energy-density is an
oscillating function of time, exactly what we would expect for a
standing wave, but without any spatial dependence.  Roughly speaking,
this is because, for every member of the statistical ensemble, there
is another member, spatially displaced by an arbitrary amount.

It is useful to compare the result (\ref{e:infomega}) with the
result normally quoted for the spectrum $\Omega_{\rm gw}(f)$ produced in
inflation (which is sometimes derived by assuming stationarity)
\cite{LesHouches}:
\bea
\left[ \Omega_{\rm gw}(f) \right]_{\text{Stationary}}& = &
{f \over \rho_c} {d \rho_{gw} \over df} \cr
& = & {16 \over 9}
{\rho_{ds} \over \rho_P} (1 + Z_{eq})^{-1}.
\label{e:omegast}
\eea
It is easy to see that in any practical experiment, one can not
discriminate between these two possibilities: Eqs.~(\ref{e:infomega})
and (\ref{e:omegast}).  The reason is simple.  In a practical
experiment lasting (say) one year, the smallest frequency resolution $df$
that can be attained is of order $df \approx 1/\text{year} \approx
10^{-8}$~Hz.  This means that the energy density (\ref{e:infomega}) is
resolvable in frequency bins not smaller than $df$. Hence the outcome
of a (noise free) experiment would be a measure of $\rho_{gw}$
averaged over a range of $k$ for which $dk = 4 \pi df/\eta_0 H_0$.
Even over this tiny range of frequency $df$, the argument of the
squared cosine passes through a range $d(k\eta)= 4 \pi df/ H_0 \sim
10^{10}$. The consequence is that in any practical experiment, the
$\cos^2(k\eta)$ is averaged over approximately $10^{10}$ cycles, after
which it is indistinguishable from $1/2$.  Thus the stationary and
non-stationary cases can not in practice be distinguished: the
inflationary prediction of non-stationarity can not be falsified.

Let us now return to the main line of reasoning and continue 
the calculation of the two-point correlation function and see if a 
two-point correlation experiment may be able to distinguish between 
the stationary and non-stationary backgrounds. For this purpose
it is helpful to compare the
inflationary model to a fictitious model Universe in which the
particles were {\it not} created in perfectly correlated pairs but
were instead formed by a stationary random process which did not
correlate particles moving in opposite directions.  In this case the
wavefunctions have a time-dependent phase and are given
by\footnote{These are not normalized modes, since they do not
correspond to vacuum fluctuations.  Instead, they are modes containing
the same energy density as in the inflationary case, but without the
correlations between the oppositely-moving quanta.}
\begin{equation}
\left[ \phi(\eta,k) \right]_{\text{Stationary}} \approx  {1 \over \sqrt{2}} ~
4 \imath k^{-5/2} \sqrt{
{8 \over 3 \pi} {\rho_{ds} \over
\rho_P} } \eta_2 \eta_0^{-2} {\rm e}^{\imath k\eta}.
\end{equation}
These ``modes'' should be compared with the ones given in
Eq.~(\ref{e:finalmode}).  They lead to same average energy density in
gravitational waves as in the squeezed-state case.

\begin{figure}
\centering
\psfig{file=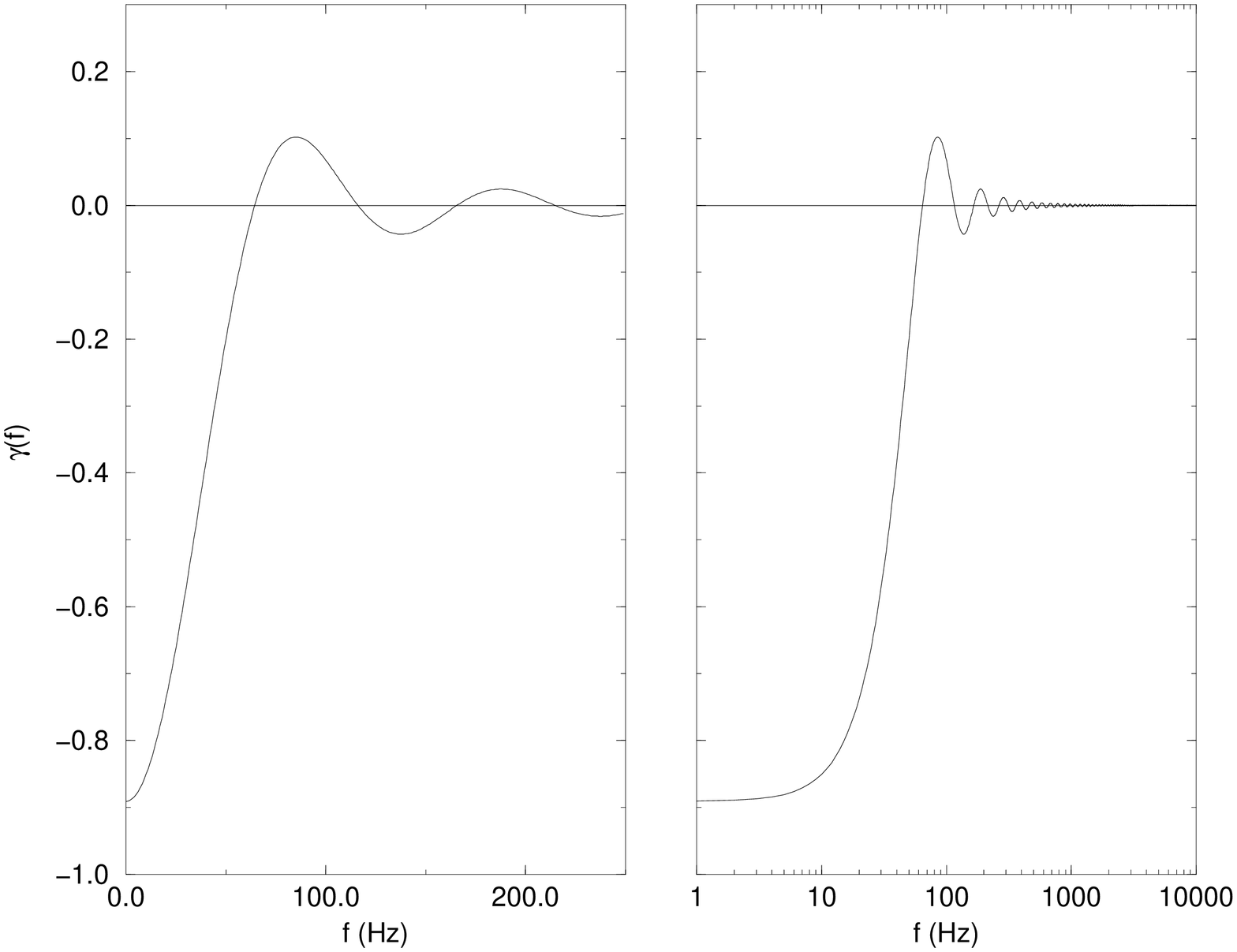,width=7.0cm,angle=-90}
\caption{The overlap reduction function $\gamma(f)$ for the
LIGO-Hanford and LIGO-Livingston sites.  The left graph has a linear
frequency scale, the right graph has a logarithmic scale.}
\label{f:overlap}
\end{figure}

The two-detector correlation
function $C(t,t')$ for the inflationary model can now be derived simply by 
substituting the
two-point function of the field operator (\ref{e:twopointfn}) into the
correlation function (\ref{e:corrfun}):
\begin{eqnarray}
&& C(t,t') = {8 \pi \over 5}  \left[ 2 \pi a(\eta_0) \right]^3 
\nonumber \\
\mbox{} && \times 
{\rm Re} \left[ \int_0^\infty df ~ \gamma(f)  f^2 \phi[\eta(t),k(f)] 
\phi^*[\eta'(t'),k(f)] \right]
\end{eqnarray}
where the overlap reduction function $\gamma(f)$ is a real function
determined entirely by the relative separation and orientation of the
two detector sites and is defined by Eq.~(3.30) of reference
\cite{allenromano} (this function was originally defined and computed in reference
\cite{Flan93}).   The overlap function $\gamma(f)$ for the two LIGO
sites is shown in Fig.~\ref{f:overlap}.  The function $\gamma(f)$
is unity for coincident and co-aligned detectors.
\noindent
Inserting the mode functions (\ref{e:finalmode}) yields
\footnote{The apparent divergence of this integral as $f \to 0$ is due
to the approximation made in its derivation that the frequency $f$ is
between $\approx 1$ mHz and $\approx 1$ kHz. The exact expression is
free of infra-red divergences, but would give the same function of $t$
and $t'$ for any practical experiment, since the measured correlation
function is given by the integral restricted to the bandpass of the
detector.}
\begin{eqnarray}
\label{e:cfun1}
C(t,t')  & & =   {8 \over 15 \pi^2}  (1+Z_{eq})^{-1} {\rho_{ds} \over \rho_P}
H_0^2 \cr
\int  df ~ \gamma(f) & &  f^{-3} \big[ \cos 2\pi f(t-t') + \cos 2\pi
f(t+t'+4 T_0) \big],
\end{eqnarray}
where we have assumed that $t$ and $t'$ are in the present epoch,
replacing $k (\eta-\eta')$ by $2\pi f(t-t')$ and $k (\eta+\eta')$ by
$2\pi f(t+t' + 4T_0)$. As mentioned earlier, the number of cycles of a wave during the infinitesimal time interval 
$dt$ at time $t$ is $dN=f(t)dt=kd\eta/ 2\pi$. Hence,   
provided that $\eta$
and $\eta'$ are not too far apart (in the cosmological sense) then one
has $k(\eta -\eta') \approx 2 \pi f (t-t')$.  This holds provided that
$|t-t'| << H_0^{-1}$.  Note that, while it is tempting to replace $k
\eta$ with $2 \pi f t$, it is incorrect.  In fact, provided that the
inflationary phase is not too long, one has (today) $k\eta \approx
2\pi f(t+ 2 T_0)$, where $T_0 = 2/3H_0$ is the present cosmological
age.  The ``additional cycles'' arise because as one goes towards the
past, the frequency of the wave increases due to blueshifting, so, for
example $k(\eta+\eta') \approx 2 \pi f (t+t' + 4 T_0)$.

In the stationary case,
which lacks correlation between the amplitudes of the
opposite-momentum modes, one would find that the correlation function
is given by
\begin{eqnarray}
\label{e:cfun2}
C(t,t') & = & {8 \over 15 \pi^2} (1+Z_{eq})^{-1} {\rho_{ds} \over \rho_P}
H_0^2 \cr
& & \int  df ~ \gamma(f)   f^{-3} \cos 2\pi f(t-t').
\end{eqnarray}
This is identical to the inflationary case, except that the
non-stationary term depending upon $t+t'$ is absent.  These
expressions are valid provided that $|t - T_0|$ and $|t'-T_0|$ are
both very small compared to $T_0$.

In a more complicated model of the early Universe, where the
energy-density during the inflationary epoch was not exactly constant
as here, but was instead a slowly varying function of time, one would
obtain an almost-identical result.  The only difference is that one
would find an extra, slowly-varying power law factor of $f^{\alpha}$
in $\Omega_{\rm gw}(f)$, and in the integrand of $C(t,t')$, where $\alpha$
is the so-called ``tilt'' of the spectrum.  This slowly-varying factor
would have no effect on our arguments or conclusions.

\subsection{Are the non-stationary terms observable?}

It is now easy to answer the original question: could a correlation
experiment carried out with two interferometric gravitational wave
detectors distinguish between the non-stationary squeezed-state
stochastic background produced by inflation, and a stationary
background with the same average energy density?  The answer is ``no''.
The reason is the same as that given following Eq.~(\ref{e:omegast}).
Suppose that there were no significant detector noise to contend with,
and that we were only trying to distinguish between the two possible
correlation functions (\ref{e:cfun1}) and (\ref{e:cfun2}).  In an
experiment of realistic length (say, one year) the smallest range of
frequencies that would be observable is a bandwidth $df = 10^{-8}$~Hz.
Consider now the extra integral term $\int df ~ \gamma(f) f^{-3} \cos
2\pi f(t+t'+4 T_0) $ which distinguishes the two cases.  The time
$t+t'+4T_0$ that appears in this integral is approximately six times
the total age of the Universe, in other words, about $10^{18}$
seconds.  Thus, even if the range of integration is restricted to
$10^{-8}$~Hz, the cosine factor appearing in this integral undergoes
more than $10^9$ cycles.  Since all of the {\it other} factors in the
integral are smoothly varying over the range of integration, the
resulting integral vanishes, in comparison with the integral
containing the stationary contribution $\cos 2\pi f(t-t')$
\footnote{ 
Note that in optimal filtering schemes to search for a stochastic
background, one is mostly concerned with times $t$ and $t'$ for which
$|t-t'| \lesssim 50~\rm ms$.}.

The same conclusion is reached when one uses the results and notations
of Ref.~\cite{grish}.  The strong dependence on frequency of the
integrand appearing in Eq.\ (14) 
of Ref.~\cite{grish}
is due to the $\cos^2 [2 \pi \nu (t - t_\nu)]$ factor, and not on the
slowly decaying $\nu^{2 \beta +1 }$ factor.  [The $2 \pi \nu t$ of
Ref.~\cite{grish} corresponds to $k \eta$ in our notation and $2\pi
\nu t_\nu$ to $k \eta_1$]. We have already shown that $2 \pi \nu t_\nu
= k \eta_1 << 1$.  However in the other term $t$ is the cosmological
age of the Universe, and $2 \pi \nu t = k \eta_0 > 10^{16}$ is ($2
\pi$ times) the number of cycles that the wave has undergone since the
beginning of the Universe. 
Thus one cannot approximate the integral by
taking its value at the lower limit of frequency, which is precisely
the approximation that appears to make the non-stationary nature of
the process visible even in year-long experiments in Ref.\
\cite{grish}. 

Finally we note that even if it were magically possible to observe the
gravitational wave correlation over a sufficiently narrow bandwidth
for the second integral to contribute significantly, the best that one
could possibly do with optimal filtering is to add a factor of 2 to
the total correlation.  This means that, even if the optimal filtering
strategy for this non-stationary signal could be implemented, the most
one would add to the total signal would be a factor of two. The total
noise would remain the same.  Hence, even if there were no problems
related to the short observation time (compared to the age of the
Universe) it would not be possible to claim that the signal-to-noise
ratio grows faster as a function of integration time for the squeezed
background than for the standard one.

\acknowledgements
BA thanks Leonid Grishchuk for many useful and stimulating discussions
about these and related topics, and Kip Thorne for some wise counsel.
MAP thanks the Caltech LIGO project for its gracious hospitality while
this paper was being written.  BA and MAP also acknowledge valuable
advice from Riccardo DeSalvo.  EF thanks the Institute for Theoretical
Physics in Santa Barbara for its hospitality, and acknowledges the
support of the Alfred P. Sloan foundation.  This work has been
supported by NSF grants PHY9728704, PHY9507740, PHY 9722189, and PHY
9407194.

\appendix

\section{Gravitational wave predictions of inflationary models}
\label{s:appendix_inflation}

In this Appendix we review the predictions \cite{grish89,starobinsky}
of inflationary models for the statistical properties of relic
gravitational waves.  We write the spacetime metric as
\be
ds^2 = a(\eta)^2 \left\{ - d\eta^2 + \left[\delta_{ab} + h_{ab}(\eta,{\bf
x})\right] dx^a dx^b \right\},
\ee
where $a,b$ run over spatial indices, $\eta$ is conformal time, and
$a(\eta)$ is the scale factor.  For the gravitational wave modes
relevant to ground and space based detectors, it is a good
approximation to take $a(\eta)$ to be constant at sufficiently late and
sufficiently early times:
\begin{equation}
\label{approx0}
a(\eta) = \left\{ \begin{array}{ll} 
 a_i
 & \mbox{\ \ \  $\eta \le \eta_i$,}\nonumber\\ 
a_f 
& \mbox{\ \ \ $ \eta > \eta_f$,}\\
\end{array} \right.
\end{equation}
for some $\eta_i$ and $\eta_f$.  For simplicity and without loss of
generality we will take $a_f = 1$.  
We define $\alpha_k$, $\beta_k$ to be the Bogoliubov coefficients
for the differential equation 
\be
\mu^{\prime\prime}(\eta) + \left( k^2 - {a^{\prime\prime} \over a}
\right) \mu(\eta) =0.
\ee
In other words, $\alpha_k$ and $\beta_k$ are such that if $\mu(\eta) =
\exp(i k \eta)$ for $\eta \le \eta_i$, then $\mu(\eta) = \alpha_k
\exp(i k \eta) + \beta_k \exp(- i k \eta)$ for $\eta \ge \eta_f$.

The metric perturbation $h_{ab}$ for $\eta > \eta_f$ can be expanded as
\be
h_{ab}(\eta,{\bf x}) = \int_{-\infty}^\infty df e^{- 2 \pi i f \eta} 
\int d^2 \Omega_n
\,\sum_{A} \,s_{A,{\bf n}}(f) \, e^{2 \pi i f {\bf n} \cdot
{\bf x}} \, e_{ab}^{A,{\bf n}}.
\end{equation}
Here $\int d^2 \Omega_n$ denotes the integral
over solid angles parametrized by the unit vector ${\bf n}$, $A$ runs
over the two polarization components,
and the tensors
${\bf e}^{A,{\bf n}}$ are the usual transverse traceless polarization
tensors, normalized according to $\delta^{ac} \delta^{bd} e_{ab}^{A,{\bf n}} e_{cd}^{B,{\bf
n}} = 2 \delta_{AB}$.  We specialize to a circular
polarization basis for which 
${\bf e}^{A,- {\bf n}} = \left({\bf e}^{A, {\bf n}} \right)^*$.
The quantities $s_{A,{\bf n}}(f)$ are Gaussian random processes 
with $s_{A,{\bf n}}(-f) = s_{A,{\bf n}}(f)^*$ and whose
statistical properties are given by \cite{note1,note2}
\begin{eqnarray}
\langle s_{A,{\bf n}}(f) && s_{B,{\bf m}}(f^\prime)^* \rangle = 
{2\over \pi} |k|  \, | \beta_k |^2 \ \delta_{AB} \ \delta(f - f^\prime) \ 
\delta^2({\bf n},{\bf m}) \nonumber \\
\mbox{} && + 
{2 \over \pi} |k| \, \alpha_k^*  \beta_k \ \delta_{AB} \ \delta(f + f^\prime)
\ \delta^2({\bf n},-{\bf m}),
\label{summary}
\end{eqnarray}
where $k = 2 \pi f$ and $\delta^2({\bf n},{\bf m})$ is the delta
function on the unit sphere.  If one drops the second term in Eq.\
(\ref{summary}), one obtains a 
stationary, Gaussian stochastic background with 
\be
\Omega_{\rm gw}(f) = {k^4 \over \pi^2 \rho_{\rm c}} \ | \beta_k|^2.
\ee
Here $\Omega_{\rm gw}(f)$ is the usual energy density per logarithmic
frequency in units of the closure energy density $\rho_c$.  The
second term in Eq.\ (\ref{summary}) encapsulates the non-stationarity.

The stochastic background will be dominated by modes for which
the number of quanta created per mode $|\beta_k|^2$ is large compared
to unity, which for typical inflation models means all modes with
frequencies $f$ in the range $10^{-18} \, {\rm Hz} \alt f \alt {\rm
MHz}$.  For such modes the ratio of the coefficients of the first
and second terms in Eq.\ (\ref{summary}) is very nearly a pure phase,
since $|\alpha_k|^2 - |\beta_k|^2=1$.
So we can write
\begin{eqnarray}
\langle s_{A,{\bf n}}(f) && s_{B,{\bf m}}(f^\prime)^* \rangle = 
{\rho_c \Omega_{\rm gw}(|f|) \over 4 \pi^2 |f|^3} \delta_{AB} \ 
\bigg[ \delta(f - f^\prime) \ 
\delta^2({\bf n},{\bf m}) 
\nonumber \\
\mbox{} && + e^{i \chi(k)}  \ \delta(f + f^\prime)
\ \delta^2({\bf n},-{\bf m}) \bigg],
\label{summary1}
\end{eqnarray}
where as before $k = 2 \pi f$ and the phase $\chi(k)$ is given by
$\exp[i \chi(k)] = \alpha_k^* / \beta_k^*$.   Hence, each inflationary
model is characterized, in the large squeezing limit, by two functions
of frequency: the spectrum $\Omega_{\rm gw}(f)$ and the phase
$\chi(k)$.  Note that under changes $\eta \to \eta - \Delta \eta$ in
the origin of conformal time, $\chi(k)$ transforms as $\chi(k) \to
\chi(k) + 2 k \Delta \eta$.  Also the phase $\chi(k)$ satisfies
$\chi(-k) = - \chi(k)$. 

In Sec.\ \ref{s:inflation} above we show that in a specific inflationary model,
$\chi(k) \approx 2 k \eta_1$, where $\eta_1$ is a specific value of
conformal time around the inflationary epoch, and that $k \eta_1 \ll
1$ for modes that are relevant for ground and space based detectors.
Thus we may take $\chi(k) \approx 0$.  This conclusion is valid for
all inflationary models.  If one uses the method of calculation of
Ref.\ \cite{grish0} to approximately evaluate $\alpha_k$ and $\beta_k$, one
finds that, up to additive corrections of order unity,
\be
\chi(k) \approx 2 k \, \eta_{k,E}
\label{generalphaseans}
\ee
where $\eta_{k,E}$ is the conformal time at which modes
with wavenumber $k$ re-enter the horizon, at which point
\be
k = a^\prime/a.
\label{horizoncrossing}
\ee
For the relevant modes which re-enter during the radiation dominated era,
$\eta_{k,E}$ is independent of the details of the inflationary
dynamics, and from Eq.\ (\ref{horizoncrossing}) is given by 
$\eta_{k,E} = \eta_0 + 1/k$, where $\eta_0$ 
is the conformal time such that $a(\eta) \propto \eta - \eta_0$ during
radiation domination (i.e. the extrapolated zero crossing of
$a(\eta)$).  Thus from Eq.\ (\ref{generalphaseans}) we find that
$\chi(k) \approx 2 + 2 k \eta_0$.  We can neglect the constant first
term, and we can choose the origin of conformal time so that
$\eta_0=0$, thus giving $\chi(k) =0$.
This conclusion is used in Sec.\ \ref{s:simplified} above, where we
use a discretized version of Eq.\ (\ref{summary1}) with $\chi(k)$ set
to zero.

\section{Detectability of squeezing in a simple model}
\label{s:appendix_stats}

In this Appendix, we analyze the detectability of the non-stationarity
in the context of the simple model of Sec. \ref{s:simplified}.  Our
starting points are Eq.\ (\ref{stats2}), which describes the statistical
properties of the Universe modes, and Eqs.\ (\ref{dft_rel}) and (\ref{wjp0}),
which describe the relationship between the Universe modes and the
measured modes.

We can summarize the information in Eq.\
(\ref{stats2}) in terms of a characteristic function:
\be
\left< \exp \left[ i \sum_{j=0}^{N-1} ( v_j {\tilde h}_j + v_j^*
{\tilde h}_j^* ) \right] \right> = \exp \left[ - {1 \over 2} \Theta(v_j)
\right],
\label{char}
\ee
where
\be
\Theta(v_j) = \sum_{j=0}^{N-1} \sigma_j^2 \left[ 4 
|v_j|^2 +  2 \varepsilon \, (v_j^2 + v_j^{*\,2} ) \right],
\label{Thetadef}
\ee
and $v_1, \ldots, v_N$ are arbitrary complex numbers.
As before, $\varepsilon=0$ is the stationary case and $\varepsilon=1$
is the squeezed case.
We can derive the corresponding characteristic function for the
measured modes ${\tilde H}_A$ by using Eqs.\ (\ref{dft_rel}) and
(\ref{char}), which yields 
\be
\left< \exp \left[ i \sum_{A=0}^{M-1} ( s_A {\tilde H}_A + s_A^*
{\tilde H}_A^* ) \right] \right> = \exp \left[ - {1 \over 2} \Theta(v_j)
\right],
\label{char1}
\ee
where now $s_1, \ldots, s_M$ are arbitrary complex numbers and
where on the RHS $v_j$ is given by 
\be
v_j = W_{Aj} s_A.
\label{whichvs}
\ee
Here and below it is assumed that repeated lower case indices $j,k,
\ldots$ are summed over $0,1,\ldots, N-1$, and upper case indices $A, B,
\ldots$ are summed over $0,1, \ldots M-1$ (see Sec.\
\ref{s:observational} above).
Now combining 
Eqs.\ (\ref{Thetadef}) and (\ref{whichvs}) yields 
\begin{eqnarray}
\Theta &=&  4 \, \Gamma_{AB} s_A s_B^*
+ 2 \varepsilon \left[ \chi_{AB} s_A s_B + \chi_{AB} s_A^* s_B^* \right],
\label{Thetaformula}
\end{eqnarray}
where
\be
\Gamma_{AB} = \sum_k \, \sigma_k^2 \, W_{Ak} W_{Bk}^*
\label{Gammadef}
\ee
is a Hermitian matrix, and 
\be
\chi_{AB} = \sum_k \, \sigma_k^2 \, W_{Ak} W_{Bk}
\label{chidef}
\ee
is a symmetric matrix.

The difference between the stationary and squeezed cases is due to the
matrix $\chi_{AB}$.  It follows from the formulas (\ref{char1}) and
(\ref{Thetaformula}) that when $\chi_{AB}=0$, the joint probability
distribution for the variables ${\tilde H}_1, \ldots, {\tilde H}_M$
is exactly the same in the squeezed and stationary cases; in
particular the phase of each ${\tilde H}_A$ is uniformly distributed
over the circle.

We now show that for a purely white stochastic background with
$\sigma_j = $  constant, the matrix $\chi_{AB}=0$ and thus the
squeezing has {\it no effect} on the statistical properties of the
measured modes ${\tilde H}_A$, irrespective of the values of $T_s$ and
$T_{\rm obs}$.
First, using the formula (\ref{wjp0}) for the weights $W_{Aj}$
we can obtain the formula for $\Gamma_{AB}$:
\begin{eqnarray}
\Gamma_{AB} &=& {1 \over N^2} \sum_{C,D} e^{2 \pi \imath (AC - BD)/M }
\sum_{j=0}^{N-1} \sigma_j^2 
\nonumber \\ \mbox{} && \times 
\ e^{-2 \pi \imath (C - D) j/N},
\label{Gammaans}
\end{eqnarray}
which can be approximately evaluated to yield
\be
\Gamma_{AB} \approx {M \over N} \, \delta_{AB} \ \sigma^2_{j = {N \over M}
A}.
\label{Gammaans1}
\ee
For the matrix $\chi_{AB}$ we find from Eqs.\ (\ref{wjp0}) and
(\ref{chidef}) the formula  
\begin{eqnarray}
\chi_{AB} &=& {1 \over N^2} \sum_{C,D} e^{2 \pi \imath (AC + BD)/M }
\sum_{j=0}^{N-1}
\sigma_j^2 \ e^{- 4 \pi \imath j s/N} \nonumber \\
\mbox{} && \times e^{-2 \pi \imath (C + D) j/N}.
\label{chians}
\end{eqnarray}
Now in the case $\sigma_j =$ const of a white spectrum, the sum over
$j$ in the formula (\ref{chians}) can be written as
\be
\sum_{j=0}^{N-1} \alpha^j = {1 - \alpha^N \over 1 - \alpha},
\label{pp}
\ee
where
\be
\alpha = \exp \left[ - 2 \pi \imath (2 j s + C + D) / N \right].
\label{pp1}
\ee
We can assume that $s>0$, which corresponds to $T_s >0$, 
since measurements today must have $T_s$ of order the age of the
Universe \cite{note4}.  Also we can assume that $s + 2 M < N$, which
corresponds to $T_s + 2 T_{\rm obs} < L$, since we should take $L \to
\infty$ at the end of our calculation anyway.  It then follows 
from Eqs.\ (\ref{pp}) and (\ref{pp1}) that $\chi_{AB}=0$.

Thus, the effect of the squeezing on the statistical properties of the
measured modes ${\tilde H}_A$ depends on the spectrum $\sigma_j$ of
the stochastic background.  Let us now turn to the case when the
spectrum is colored.  In this case it turns out that the matrix
$\chi_{AB}$ is small, and consequently the effects of the squeezing
are small, whenever the observation starting time $T_s$ is large
compared to a correlation time that characterizes the spectrum of the
stochastic background.  In other words, the effect of the squeezing on
the statistical properties of the stochastic background is time
dependent; the effect is strong near $t=0$, where the modes are all
synchronized \cite{note4}, but becomes weaker and weaker at later
times.  

To see this, it is convenient to transform back to a continuum
representation of the stochastic background.  
The spectrum $S_h(f)$ of the background $h(t)$ is related to
the quantities $\sigma_j$ by \cite{derive}
\be
\sigma_j^2 = {1 \over 2} N^2 \Delta f S_h(j \Delta f),
\label{discrete1}
\ee
where $\Delta f = 1/ (2 \pi L)$.
We define the quantity
\be
{\hat C}_h(\tau) = \int_0^\infty df \ e^{- 2 \pi \imath f \tau} \
S_h(f).
\label{hatCdef}
\ee
Then the real part of ${\hat C}_h(\tau)$ is the usual correlation
function $\langle h(t) h(t+\tau) \rangle$ (in the stationary case).
The discrete form of Eq.\ (\ref{hatCdef}) is 
\be
{\hat C}_h(j \Delta \tau) = {2 \over N^2} \sum_{k=0}^{N-1} e^{- 2 \pi
\imath {j k \over N}} \, S_h(k \Delta f),
\label{hatCdef1}
\ee
where $\Delta f = 1 / (2 \pi L)$ and $\Delta \tau = 2 \pi L / N = L
\Delta t$.  Now by combining Eqs.\ (\ref{discrete1}) and (\ref{hatCdef1})
we can evaluate the sum over $j$ that appears in the formula (\ref{chians})
for $\chi_{AB}$:
\be
\sum_j \sigma_j^2 e^{- 2 \pi \imath (2s + C + D) j/N} = {1 \over 2}
N^2 {\hat C}_h[2 T_s + (C + D) \Delta \tau].
\label{eee}
\ee

The quantity (\ref{eee}) will be small once $T_s$ is 
much larger than the correlation time $\tau_*$ of ${\hat C}_h$, that
is, the time over which ${\hat C}_h(\tau) \sim {\hat C}_h(0)$.  This 
correlation time $\tau_*$ 
will be roughly the reciprocal of the shortest frequency scale over
which $S_h(f)$ has appreciable variation.  The ``high 
frequency'' structure in the spectrum $S_h(f)$ will presumably be
dominated by the ``breaks'' in the spectrum $S_h(f)$ corresponding
to transitions from one cosmological epoch to another (eg inflation to
radiation domination).  However, the physical time for such
transitions to take place cannot be shorter than a local Hubble time.
Hence, the corresponding correlation time $\tau_*$ will always be much
shorter than the present day age of the Universe $T_0$, and
consequently always much smaller than $T_s$ for realistic
measurements.  Hence the matrix $\chi_{AB}$ and consequently the
effect of the squeezing on observational data will be very small.

Turn, now, to the question of how accurately the matrix $\chi_{AB}$
can be measured, i.e., to the fundamental limitations on measurement
accuracy imposed by ``cosmic variance''.
To address this question, we apply the analysis of Appendix 
\ref{s:appendix_stats1} to deduce the conditions under which the
squeezing is detectable.  Note that this analysis
allows for the possibility of combining the
measurements of all the different mode amplitudes.
Lets define the real random vector ${\bf x} = ( {\rm Re} {\tilde H}_1,
\ldots, {\rm Re} {\tilde H}_M, {\rm Im} {\tilde H}_1, \ldots, {\rm Im}
{\tilde H}_M)$.  Then, 
from Eqs.\ (\ref{char1}) and (\ref{Thetaformula}), the analysis of 
Appendix \ref{s:appendix_stats1} applies directly with
\begin{equation}
{\bf \Sigma}_1 = 
4 \left[
\begin{array}{cc}
 {\rm Re} \, {\bf \Gamma} &  {\rm Im} \, {\bf \Gamma} \\
 - {\rm Im} \, {\bf \Gamma} &  {\rm Re} \, {\bf \Gamma}
\end{array}\right]
\end{equation}
and 
\begin{equation}
{\bf \Sigma}_2 - {\bf \Sigma}_1= 
4 \left[
\begin{array}{cc}
 {\rm Re} \, {\bf \chi} &  - {\rm Im} \, {\bf \chi} \\
 - {\rm Im} \, {\bf \chi} &  -{\rm Re} \, {\bf \chi}
\end{array}\right].
\end{equation}
It then follows from Eq.\ (\ref{finalans}) that the difference between
the squeezed and stationary cases is only detectable in the regime
where the quantity
\be
\Lambda = {\rm tr} \left[ 
\left( {\bf \chi} \cdot {\bf \Gamma}^{-1}
\right) \ 
\left( {\bf \chi} \cdot {\bf \Gamma}^{-1}
\right)^\dagger \right]
\label{Lambdadef1}
\ee 
is large compared to unity.
We can approximately evaluate the quantity $\Lambda$ by 
substituting Eqs.\ (\ref{Gammaans1}), (\ref{chians}), and 
(\ref{eee}) into Eq.\ (\ref{Lambdadef1}).  This yields, in the limit 
$L \to \infty$ and $\Delta t \to 0$,
\be
\Lambda \sim \sum_{j=0}^\infty { \left| \int_{0}^{2 T_{\rm
obs}} d \tau \ e^{2 \pi \imath j \tau / T_{\rm obs}} \ {\hat C}_h(T_s
+ \tau)
\right|^2 \over S_h(j/T_{\rm obs})^2}.
\label{ans1}
\ee

We now analyze the implications of the final result (\ref{ans1}).  
First, it follows from the formula (\ref{ans1}) together with the
definition (\ref{hatCdef}) that in the regime $T_{\rm obs} \ll T_s$,
we always have $\Lambda \alt 1$, irrespective of the nature of the
spectrum $S_h(f)$, which proves the claims made in the body of the
paper.  In the other regime where $T_{\rm obs} \sim T_s$, 
we can see that $\Lambda \ll 1$ once $T_s$ is
much larger than the correlation time of $\tau_*$ of ${\hat C}_h$.
As argued above, $\tau_*$ will be much smaller than the present day
age of the Universe $T_0$.  Hence, 
experiments today would have difficulty detecting the
non-stationarity even if they last a time $\sim T_0$.

\section{Distinguishing between two different distributions of a
Gaussian random vector} 
\label{s:appendix_stats1}

In this Appendix we address the following statistical issue, which 
arises in Sec.\ \ref{s:observational} and in Appendix
\ref{s:appendix_stats} above.  Suppose that 
${\bf x} = (x_1, \ldots, x_N)$ 
is a zero-mean, Gaussian random vector which satisfies
either
\be
\langle x_i x_j \rangle = \Sigma_{1\ ij}
\label{case1}
\ee
(case 1) or 
\be
\langle x_i x_j \rangle = \Sigma_{2\ ij}
\label{case2}
\ee
(case 2).  Thus, there are two possible variance-covariance matrices, ${\bf
\Sigma}_1$ and ${\bf \Sigma}_2$.  
In our application we will take case 2 to correspond to a
squeezed stochastic background, and case 1 to a stationary stochastic
background.  Suppose now that we have one measurement
of ${\bf x}$.  How well can we distinguish between the two
possibilities?

We now show that, when ${\bf \Sigma}_1 - {\bf \Sigma}_2
\ll {\bf \Sigma}_1$, and when case 2 actually applies, the two cases
can be distinguished with high probability only in the regime where
the quantity 
\be
{1 \over 2} {\rm tr} \, \left[ \left( {\bf \Sigma}_2 \cdot
{\bf \Sigma}_1^{-1} -{\bf 1} \right)^2 \right]
\ee
is large compared to unity, and not when this quantity is of order
unity.

It is easiest to address the question using the Bayesian approach.
Let the experimenter's prior probability for case 2 be $p_2$.  Then
after the measurement her probability for case 2 will be revised to
$p_2^\prime$, where
\be
{p_2^\prime \over 1 - p_2^\prime} = {p_2 \over 1 - p_2} \ e^\Lambda,
\ee
where 
\be
e^\Lambda = {p({\bf x}|2) \over p({\bf x}|1)}
\label{Lambdadef}
\ee
and $p({\bf x}|1)$ is the probability that ${\bf x}$ is observed
assuming case 1, etc.  It is clear that the difference between cases 1
and 2 is detectable in the regime $\Lambda \gg 1$, but not in the
regime $\Lambda \sim 1$.

Inserting Gaussian probability distributions into Eq.\ (\ref{Lambdadef}) 
we find that
\be
\Lambda = {1 \over 2} {\bf x}^T \cdot \left[ {\bf \Sigma}_1^{-1} -
{\bf \Sigma}_2^{-1} \right] \cdot {\bf x} 
+ {1 \over 2} \ln {\rm det} \, ( {\bf \Sigma}_1 \cdot {\bf \Sigma}_2^{-1}).
\ee
Lets now assume that case 2 actually applies, so that the expected
value of $x_i x_j$ is given by Eq.\ (\ref{case2}).  Then, typical
values of $\Lambda$ will be close to the expected
value of the $\Lambda$ which is
\be
\langle \Lambda \rangle = - {1 \over 2} {\rm tr} \, \left[ {\bf 1} -
{\bf \Sigma}_2 \cdot {\bf \Sigma}_1^{-1} \right] 
- {1 \over 2} \ln {\rm det} \, ( {\bf \Sigma}_2 \cdot {\bf
\Sigma}_1^{-1}).
\label{exactans}
\ee
Now suppose that the eigenvalues of ${\bf \Sigma}_2 \cdot {\bf
\Sigma}_1^{-1}$ are $1 + \lambda_j$ for $1 \le j \le N$.  This yields
that
\be
\langle \Lambda \rangle = {1 \over 2} \sum_{j=1}^N \left[ \lambda_j -
\ln (1 + \lambda_j) \right].
\ee
If each $|\lambda_j|$ is small compared to unity, then to a good
approximation we have
\begin{eqnarray}
\langle \Lambda \rangle &\approx& {1 \over 4} \sum_{j=1}^N \ \lambda_j^2
\nonumber \\
\mbox{} &=& {1 \over 2} {\rm tr} \, \left[ \left( {\bf \Sigma}_2 \cdot
{\bf \Sigma}_1^{-1} -{\bf 1} \right)^2 \right].
\label{finalans}
\end{eqnarray}
In our application we use the formula (\ref{finalans}) which should be
a good approximation to the exact formula (\ref{exactans}).


\begin{thebibliography}{11}

\bibitem{KolbTurner}
Edward W. Kolb and Michael S. Turner, {\it The
Early Universe}, Addison-Wesley: Redwood City, California (1990).

\bibitem{Linde}
Linde, A. D., {\it Particle Physics and Inflationary
Cosmology}, Harwood, Chur, Switzerland, 1990.

\bibitem{RecentReview}
M. Turner, {\it Cosmological Parameters}, astro-ph/9904051, to be
published in {\it The Proceedings of Particle Physics and the Universe
(Cosmo-98)}, edited by David O.  Caldwell (AIP, Woodbury, NY).

\bibitem{StringReview}
R.A. Battye, {\it Cosmic strings in a universe with non-critical matter
density}, astro-ph/9806115 to appear in {\it Fundamental Parameters in
Cosmology}, Recontres de Moriond, Jan. 1998.

\bibitem{LesHouches}
B.~Allen, in {\em Proceedings of the Les Houches
School on Astrophysical Sources of Gravitational Waves},
eds. J.A.~Marck and J.P.~Lasota, Cambridge, 373 (1997).

\bibitem{allenromano}
B. Allen and J.D. Romano, {\it Detecting a stochastic background of
gravitational radiation: signal processing strategies and
sensitivities}, (1997) gr-qc/9710117, to appear in Physical Review D (1999).

\bibitem{ligo}
A.~Abramovici {\it et al}., Science {\bf 256}, 325 (1992).

\bibitem{virgo}
B.~Caron {\it et al}., in {\it Gravitational Wave Experiments}, edited by
E. Coccia, G. Pizzella, and F. Ronga, (World Scientific, Singapore, 1995).

\bibitem{geo}
K.~Danzmann et al., in {\it Gravitational Wave Experiments}, edited by
E. Coccia, G. Pizzella, and F. Ronga, (World Scientific, Singapore,
1995).

\bibitem{lisa}
A good general review of the LISA project is the {\it
Proceedings of the Second International LISA Symposium}, Pasadena, July
1998 (ed. W. Folkner, American Institute of Physics).

\bibitem{grish}
L.P. Grishchuk, {\it The detectability of relic (squeezed)
gravitational waves by laser interferometers}, gr-qc/9810055.

\bibitem{grish89}
L.P. Grishchuk and Yu. V. Sidorov, Class. Quant. Grav. {\bf 6}, L161 (1989).

\bibitem{grishnasa}
L.P. Grishchuk, in {\it Workshop on squeezed states and uncertainty
relations}, NASA Conf. Publ. {\bf 3135}, 1992, p. 329; Class. Quant.
Grav. {\bf 10}, 2449 (1993).

\bibitem{starobinsky}
D. Polarski, A. A. Starobinsky, Class. Quant. Grav. {\bf 13}, 377
(1996) (also gr-qc/9504030).  A similar discussion but in the context
of scalar rather than tensor perturbations is given in Ref.\
\cite{albrecht0}.  

\bibitem{albrecht0}
A. Albrecht, P. Ferreira, M. Joyce, and T. Prokopec, Phys. Rev. D {\bf
50}, 4807 (1994) (also astro-ph/9303001). 

\bibitem{sqrt}
Reference \cite{grish} refers to the signal-to-noise ratio (SNR)
growing as (or faster than) than the fourth-root of the observation
time, because there, the signal is an amplitude.  This is equivalent to
our discussion, where the signal ($C(t,t')$ or some filtered version of
it) is an amplitude-squared (a power) proportional to $\Omega_{\rm gw}$,
growing as (or faster than) the square root of the observation time.
See for example Eq.~(3.33) of Ref~\cite{LesHouches} or Eq.~(3.75) of
Ref.~\cite{allenromano}.

\bibitem{albrecht1}
A. Albrecht, {\it Coherence and Sakharov Oscillations in the Microwave
Sky}, astro-ph/9612015.

\bibitem{KK}
M. Kamionkowski, A. Kosowsky, Phys. Rev. D {\bf 57}, 685 (1998).

\bibitem{starobinskyaa}
J. Lesgourgues, D. Polarski, S. Prunet, and A. A. Starobinsky,
{\it Detectability of the primordial origin of the gravitational
wave background in the Universe}, gr-qc/9906098, June 1999.

\bibitem{LIGOeg}
An example that illustrates
the necessity of such an analysis is the hoped-for detection of
coalescing compact binaries with LIGO, where the overall signal is
detectable, but where nevertheless the signal in each individual time
bin of the data is small compared to the noise in that bin.

\bibitem{turnergw}
Michael S. Turner, Phys. Rev. {\bf D55}, 435 (1997).

\bibitem{AK94}
B. Allen and S. Koranda, Phys. Rev. {\bf D50}, 3713 (1994).

\bibitem{explain}
The reason for this is as follows.
The discussion of the previous paragraph applies only to modes of fields
for which geometric optics is valid.  Modes which ``leave the
horizon'' during inflation 
violate this assumption, and are parametrically amplified.  Now those 
modes which are inside the horizon today and which were subject to
parametric amplification (i.e., at some stage left
the horizon) had physical wavelengths $\lambda_{\rm phys}$ at the
beginning of inflation in the range 
$$
e^{-N} H_i^{-1} \alt \lambda_{\rm phys} \alt Z_{\rm end} e^{-N} H_i^{-1} /
\sqrt{Z_{\rm eq}},
$$
where $H_i$ is the Hubble constant during inflation, $N$ is the number
of efoldings during inflation, $Z_{\rm eq}$ is the redshift of
matter-radiation equality, and $Z_{\rm end}$ is the redshift of the end of
inflation.  From the parameter values in Sec.\ \ref{twopoint},
we see that these wavelengths are so tiny that it is natural to
assume that the corresponding modes all started in their vacuum states
at the beginning of inflation.  This is the assumption that is usually
made.  

\bibitem{jurassic_era}
L.\ P.\ Grishchuk, Zh. Eksp. Teor. Fiz. {\bf 67}, 825 (1974)
[Engl. transl. in Sov. Phys. JETP {\bf 40}, 409 (1975)].  

\bibitem{allen89}
See the discussion and references in
B. Allen, Phys. Rev. {\bf D37} (1988) 2078.

\bibitem{birreldavies}
N.D. Birrell and P.C.W. Davies, {\it Quantum Fields
in Curved Space} (Cambridge University Press,
Cambridge, England, 1982).

\bibitem{parker}
L. Parker, Phys Rev. {\bf 183}, 1057 (1969), particularly the
discussion following Eq.~(50).  This early paper incorrectly states
that there is no particle production for a massless field - this was
corrected in later work

\bibitem{parker2}
L. Parker, Phys Rev {\bf D12}, 1519 (1975), particularly the discussion
between Eqs.~(34) and (40).

\bibitem{Flan93}
E. Flanagan, Phys. Rev. {\bf D48}, 2389 (1993).

\bibitem{note1}
Strictly speaking $s_{A,{\bf n}}(f)$ is a quantum mechanical
operator; see Sec.\ \ref{s:inflation}.  However, the commutator $
[ s_{A,{\bf n}}(f), s_{B,{\bf 
m}}(f^\prime)^\dagger]$ is smaller than the quantity (\ref{summary})
by a factor $\sim 1/|\beta_k|^2$, which is negligible in the limit of
large squeezing $|\beta_k| \gg 1$ (ensured by the large number of
efolds during inflation).  Thus it is a good approximation to treat
$s_{A,{\bf n}}(f)$ as classical random processes.

\bibitem{note2} 
Note that the corresponding equation in footnote [21] of Ref.\
\cite{Flan93} has a typo; the $\delta^2({\bf n},{\bf m})$ should be
corrected to $\delta^2({\bf n},-{\bf m})$.  Also the RHS of Eq.\ (2.8)
of Ref.\ \cite{Flan93} should be divided by $\pi$.

\bibitem{grish0}
L.\ P.\ Grishchuk and M.\ Solokhin, Phys. Rev. D {\bf 43}, 2566 (1991).
See in particular the equation before Eq.\ (18).

\bibitem{note4}
Observations that start at $t=0$ (which is of course unrealistic) can
easily distinguish between the stationary and squeezed cases, since
the initial time derivative ${\dot h}(t=0)$ is constrained to vanish
in the squeezed case, by Eqs.\ (\ref{univ_modes}) and (\ref{phaseg}).

\bibitem{derive}
This is most easily derived by using
$$
\langle h(t)^2 \rangle = \int_0^\infty df \ S_h(f) = {2 \over N^2}
\sum_j \sigma_j^2.
$$
Note also that the frequency $f$ is physical frequency and not
coordinate frequency. 
\end{thebibliography}
\end{document}